\documentclass[%
 reprint,
 amsmath,amssymb,
 aps,
pra,
]{revtex4-2}

\usepackage{braket}
\usepackage{subcaption}
\usepackage{dsfont}
\usepackage[mark= ]{sectionbreak}

\usepackage{graphicx}
\graphicspath{{images/}}
\usepackage{dcolumn}
\usepackage{bm}
\usepackage{mathtools}
\usepackage[thinc]{esdiff}
\usepackage{textcomp}
\usepackage{xcolor}
\usepackage{caption}
\begin{document}

\title{ Effects of XX-catalysts on quantum annealing spectra with perturbative crossings }

\author{Natasha Feinstein}
\email{ucapnjs@ucl.ac.uk}
\address{%
 London Centre for Nanotechnology, University College London, WC1H 0AH London, UK
}%

\author{Louis Fry-Bouriaux}
\address{%
 London Centre for Nanotechnology, University College London, WC1H 0AH London, UK
}%

\author{Sougato Bose}
\address{%
 Department of Physics and Astronomy, University College London, London WC1E 6BT, UK
}%

\author{P A Warburton}
\address{%
 London Centre for Nanotechnology, University College London, WC1H 0AH London, UK
}%
\address{%
 Department of Electronic \& Electrical Engineering, University College London, WC1E 7JE London, UK
}%

\date{\today}


\begin{abstract}
In adiabatic quantum annealing the required run-time to reach a given ground-state fidelity is dictated by the size of the minimum gap that appears between the ground and first excited state in the annealing spectrum. In general the presence of avoided level crossings demands an exponential increase in the annealing time with the system size which has consequences both for the efficiency of the algorithm and the required qubit coherence times.  One promising avenue being explored to produce more favourable gap scaling is the introduction of non-stoquastic XX-couplings in the form of a catalyst - of particular interest are catalysts which utilise accessible information about the optimisation problem in their construction. Here we show extreme sensitivity of the effect of an XX-catalyst to subtle changes in the encoding of the optimisation problem. In particular, we observe that a targeted catalyst containing a single coupling at constant strength can significantly reduce the gap closing with system size at an avoided level crossing. For slightly different encodings of the same problems however, these same catalysts result in closing gaps in the annealing spectrum. To understand the origin of these closing gaps, we study how the evolution of the ground-state vector is altered by the presence of the catalyst and find that the negative components of the ground-state vector are key to understanding the response of the gap spectrum. We also consider how and when these closing gaps could be utilised in diabatic quantum annealing protocols - a promising alternative to adiabatic quantum annealing in which transitions to higher energy levels are exploited to reduce the run time of the algorithm.
\end{abstract}

\maketitle

\section{ Introduction \label{sec:introduction} }

Quantum annealing (QA) is a continuous time quantum algorithm proposed as a means of solving NP combinatorial optimisation problems faster than can be achieved classically \cite{Apolloni1989, Finnila1994, Kadowaki1998}. Being able to solve these kinds of problems has real world applications ranging from portfolio optimisation \cite{rubiogarcía2022portfolio} and resource allocation \cite{Gabbassov_2022,witt2022tactile,otsuka2022highspeed} to transport optimisation \cite{sales2023adiabatic}. In these applications it is of great interest to be able to find the best solution in the shortest possible time.

QA proceeds by initialising a quantum system in the ground-state (GS) of a local transverse field, and evolving to a Hamiltonian whose ground-state encodes the optimal solution of the problem to be solved. If the evolution proceeds adiabatically, the ground-state of the final Hamiltonian is obtained with high probability at the end of the anneal. How fast the anneal can be carried out while keeping the system in its ground-state is determined by the adiabatic theorem \cite{TosioKato1950}. In its simplest form, this states that the total evolution duration must be inversely proportional to the square of the minimum energy gap between the ground and first excited states in the energy spectrum of the total Hamiltonian. The scaling with the problem size of the annealing time required to maintain adiabatic evolution thus depends on scaling of this gap minimum. 

Early results that focused on randomly generated instances of the exact-cover problem for small system sizes suggested that the minimum energy gap scaling may be polynomial \cite{Farhi2001}, which would allow QA to find the solution efficiently. However subsequent studies have suggested that we can generally expect exponentially closing gaps in the annealing spectrum resulting from the presence of first order phase transitions \cite{Young2010,Jorg2010,Jorg2010a,Seoane2012,Knysh2016}. A particular problem which has been highlighted is the potential for so called perturbative crossings to form between the low energy eigenstates of the problem Hamiltonian towards the end of the anneal \cite{Amin2009, Altshuler2010}. The corresponding energy gap has been found to be exponentially small with the Hamming distance between the eigenstates, which can generally be expected to grow with the system size \cite{Amin2009}. Additionally, it was argued in \cite{Altshuler2010} that the probability of such a crossing occurring would grow exponentially with the number of states that had energy comparable to that of the ground-state, suggesting that the likelihood of not encountering such a transition vanishes with increasing problem size.

Overall, these results confirm the standard assumption that quantum algorithms such as QA are unlikely to be able to change the complexity class of NP-hard optimisation problems \cite{Preskill_2018}. What is more likely is that they will offer a quantitative improvement to the run-time, even if the time scaling remains exponential with the problem size. Indeed, there have already been some promising results in this regard, with QA showing a quadratic improvement in time scaling over simulated annealing for finding the ground-state of random Ising chains \cite{Zanca2016} as well as solving the Max Independent Set (MIS) problem \cite{Ebadi2022}. However, even if we are not expecting QA to be a polynomial time algorithm, the exponential scaling in the annealing time still presents a problem in that it demands exponentially long qubit coherence times. Coherent quantum dynamics were recently demonstrated on the D-Wave Advantage hardware \cite{king2022}. Significant effects of decoherence were however seen for annealing times greater than a few nanoseconds. Other platforms may have longer coherence times than this \cite{Ebadi2022}. Without the error correcting framework that exists for gate based quantum computation, however, exponentially long annealing runs remain intractable. 

It is therefore desirable to move the exponential scaling from the anneal time to another part of the algorithm such that, even if the algorithm as a whole has an exponential total run time, the length of each annealing run is sub-exponential and therefore more likely to be within the coherence lifetime of the hardware. The most straightforward way to do this is to carry out an exponential number of anneals with sub-exponential run-time. In this way, one can still achieve a high probability of measuring the GS overall - even if the final ground-state overlap at the end of each individual anneal is exponentially small as a result of being far from the adiabatic limit. One way to improve upon this basic approach would be to find ways to utilise information obtained from each annealing run to alter the annealing path in some targeted way in order to enhance the final GS overlap of the subsequent anneals. 

A number of works exist in the literature which examine how different changes to the annealing path affect the gap spectrum, and therefore the final GS overlap. These studies typically focus on the inhomogeneous driving of the transverse field \cite{Farhi2011, Dickson2011a, Susa2018, Takada2020a, Fry-Bouriaux2021, Adame2020, Rams2016} or the introduction of an additional non-monotonic term to the Hamiltonian referred to as a catalyst \cite{Farhi2002, Seoane2012, Seki2015, Hormozi2017, Nishimori2017, Albash2019, Takada2020a, Choi2021, Albash2021, Mehta2021}. In general, the goal has been to enhance the size of the minimum gap in the annealing spectrum \cite{Farhi2011,Dickson2011a,Susa2018,Takada2020a,Adame2020,Farhi2002,Seoane2012,Hormozi2017, Nishimori2017, Albash2019,Mehta2021}. However, a higher GS fidelity can also be obtained by manipulating the spectrum such that a diabatic anneal is possible \cite{Hormozi2017, Choi2021, Fry-Bouriaux2021}. In this case the goal is to produce a path through the annealing energy spectrum that exploits transitions to higher energy states such that the system subsequently returns to the ground-state. When this can be achieved, the small energy gaps at which these transitions happen no longer present bottlenecks to the algorithm, and if all other gaps in the spectrum close no faster than polynomially with system size, a polynomial QA run will end with a high overlap with the ground-state of the problem Hamiltonian. This exploitation of transitions has been shown to result in exponential speedup for the glued trees problem whose structure results in a symmetric annealing spectrum that is naturally amenable to diabatic QA \cite{Nagaj2012}.

For the most part, these studies have not attempted to tailor the path change to the specific problem instance but rather have been studying the effectiveness of different drivers and catalysts in different settings. However there are some examples in the literature of where problem specific information is utilised. It was shown in \cite{Dickson2011a} how knowledge of the local optima could be used to adjust the relative strengths of the local X-fields to remove or weaken perturbative crossings. The same authors worked this idea into a recursive strategy that utilised the fact that sub-exponential annealing runs were likely to return the local optima responsible for the formation of the perturbative crossings in the spectrum. A more recent work \cite{Choi2021} showed that by coupling local optima to each other, one can replace a single avoided level crossing with a correlated double avoided level crossing - thus facilitating a diabatic anneal. Further to this, diagonal catalysts that bias towards the GS have been found to result in significant gap enhancement \cite{Albash2021}. However, as the authors note, knowledge of the local optima does not always give a good indication of where in the state space the global optimum might be. 

There has been a particular interest in understanding if the introduction of XX-couplings could result in more favourable gap scaling. For the p-spin \cite{Seoane2012, Nishimori2017, Albash2019} and Hopfield models \cite{Seki2015}, all-to-all non-stoquastic XX-catalysts have been shown, both analytically and numerically, to reduce the first order phase transition to second order. This is in contrast to more recent results \cite{Crosson2020a} which suggested that non-stoquastic Hamiltonians will generally have smaller gaps than stoquastic ones. Additionally, for the weak-strong cluster problem, it was found that whether stoquastic or non-stoquastic XX-couplings removed the phase transition depended on where the couplings were applied \cite{Takada2020a}. For the frustrated Ising ladder it was found that neither stoquastic nor non-stoquastic XX-interactions were able to remove the first order phase transition for their choice of couplings \cite{Takada2020}. The fact that the introduction of XX-interactions can have such strikingly different results, depending on the system they are being applied to and which specific couplings are used, suggests that it may be possible to utilise problem specific information in the construction of such catalysts. Indeed, some light has been shone on how the effect of particular XX-interactions relates to the couplings they produce between states in the problem Hamiltonian \cite{Albash2019,Choi2020}.

In this work, we aim to further add to the understanding of how different catalyst Hamiltonians affect the annealing spectrum. Specifically, we examine the effect of a targeted XX-catalyst, inspired by ideas introduced in \cite{Albash2019,Choi2020,Choi2021}, and chosen to enhance the minimum gap size at a perturbative crossing. Using the maximum weighted independent set (MWIS) problem to construct our annealing instances, we show that the effect of this catalyst is strongly dependent on arbitrary small changes to the parameters of the problem Hamiltonian which alter the properties of the perturbative crossing that forms in the original spectrum. In particular we observe that, when applied to a setting where the exponential closing of the gap minimum associated with the perturbative crossing is weaker, the catalyst is able to produce significant gap enhancement, i.e. an increase in the gap size from its value when no catalyst is applied. The catalyst strength required to maximise this gap enhancement is found to approach a constant value with increasing system size. When applied to a setting where the gap scaling associated with the perturbative crossing is stronger however, this same catalyst results in closing gaps in the annealing spectrum. More specifically, we observe that as we increase the strength of the catalyst, we first observe a closing of the gap at the perturbative crossing followed by the formation of an additional gap minimum earlier in the anneal.

We structure this paper as follows. We first introduce our annealing problem setting in Section \ref{sec:setting}. The structure of the annealing Hamiltonian and the catalyst term is described in Section \ref{sec:hamiltonian}. In Section \ref{sec:graph} we describe our process for creating MWIS problems and its relationship with the theory of perturbative crossings. In Section \ref{sec:parameters} we demonstrate how the MWIS problem parameters can be tuned to produce such a perturbative crossing, and also how they affect the scaling of the minimum energy gap against system size. With the problem setting established we then, in Section \ref{sec:results}, examine the effects of introducing targeted non-stoquastic XX-catalysts. We describe our choice of catalyst and show, in Section \ref{sec:enhanement}, that for parameter settings which result in a perturbative crossing with weaker associated gap scaling, the catalyst is able to drastically reduce the gap closing rate with system size. We then, in Section \ref{sec:closing_gaps}, apply the same catalysts to problems for which the perturbative crossing has a stronger associated gap scaling, and in this setting, we show that the effectiveness of the catalyst in enhancing the minimum gap size is drastically reduced. It is in this setting that we observe the aforementioned closing gaps in the annealing spectrum. In order to shed some light on the source of these closing gaps, as well as why the effect of the catalyst differs in these two settings, we examine how the evolution of the ground-state vector is affected by the presence of the catalyst. We also present results for intermediate parameter settings in Section \ref{sec:intermediate}. Finally, in Section \ref{sec:discussion}, we discuss the implications of our results with regards to where we expect these catalysts to be useful, how information from short annealing runs could be utilised in their construction, and also how these catalysts may be able to be extended to more complex and general graph structures. 

\section{ Problem Setting \label{sec:setting} }

We are interested in examining the effects of XX-couplings on perturbative crossings - a type of avoided level crossing (AC) that can form towards the end of the anneal between problem states that are close in energy. In particular, we wish to compare the effects of the same XX-coupling on perturbative crossings with different associated gap scaling. We construct a scalable instance of the MWIS problem that allows us to easily adjust key properties of the problem Hamiltonian that are responsible for the formation of such crossings. The MWIS problem is a natural choice due to (a) the free parameters present in the encoding of the problem into an Ising Hamiltonian and (b) the ease with which we can change the energies of the states which are one bit flip apart from each other. (The importance of this will become clear in Section \ref{sec:graph}.) The MWIS problem is also NP-complete meaning that any other problem in NP can be mapped on to it.

\subsection{ Hamiltonian \label{sec:hamiltonian} }

Our annealing Hamiltonian takes the form
\begin{equation}
    H(s) = (1-s) H_d + s(1-s) H_c + s H_p,
    \label{eq:H(s)}
\end{equation}
where $H_d$, $H_c$ and $H_p$ are independent of $s$ and denote the driver, catalyst and problem Hamiltonians respectively. The dimensionless annealing parameter, $s$, is varied from $0$ to $1$ over the course of the anneal such that $H(s)$ evolves from $H(0)=H_d$ to $H(1)=H_p$. We consider the anneal in a static setting and so it is not necessary to specify how $s$ is varied with time except to say that it increases monotonically such that the same part of the annealing spectrum is not crossed more than once. We refer to the eigenstates of the total Hamiltonian as the \textit{instantaneous} eigenstates and denote them and their corresponding energies as
\begin{equation}
    H(s) \ket{E_a(s)} = E_a(s) \ket{E_a(s)}.
    \label{eq:insts}
\end{equation}
The states are labelled starting from $a=0$ in order of increasing energy. Similarly we denote the \textit{problem} eigenstates and their energies as
\begin{equation}
    H_p \ket{E_a} = E_a \ket{E_a}.
    \label{eq:probs}
\end{equation}
Since the problem Hamiltonian is diagonal in the computational basis, the set of problem states $\{\ket{E_a}\}$ is simply the computational basis with the states labelled by energy. 

Our choice of driver is the conventional homogeneous local X-field,
\begin{equation}
    H_d = - \sum \limits_{i = 1}^{n} \sigma_i^x,
    \label{eq:Hd}
\end{equation}
where $n$ is the total number of qubits and $\sigma_x^i$ denotes the Pauli-X operator on the $i$th qubit. Its ground-state is the equal superposition over all computational basis states. The catalysts used in this work contain a single XX-coupling between two qubits and can thus be written,
\begin{equation}
    H_c = J_{xx} \sigma_i^x \sigma_j^x
    \label{eq:Hc}
\end{equation}
where $J_{xx}$ is always chosen to be positive such the total Hamiltonian is non-stoquastic for $s \neq 0,1$. Its magnitude can be tuned to adjust the strength of the catalyst relative to $H_d$ and $H_p$. There has been much discussion in the literature regarding the importance of non-stoquasticity in QA \cite{gupta2019elucidating,Choi2021,Aharonov2007}. We note that here we are using the term non-stoquastic simply to describe a Hamiltonian with positive as well as negative off diagonal elements and make no assumptions regarding the relationship between non-stoquasticity and computational complexity. 
We now go on to describe the motivation for and construction of our problem Hamiltonians. 

\subsection{ Problem Graph \label{sec:graph} }

\begin{figure*}[]
    \centering
    \includegraphics[scale=0.44]{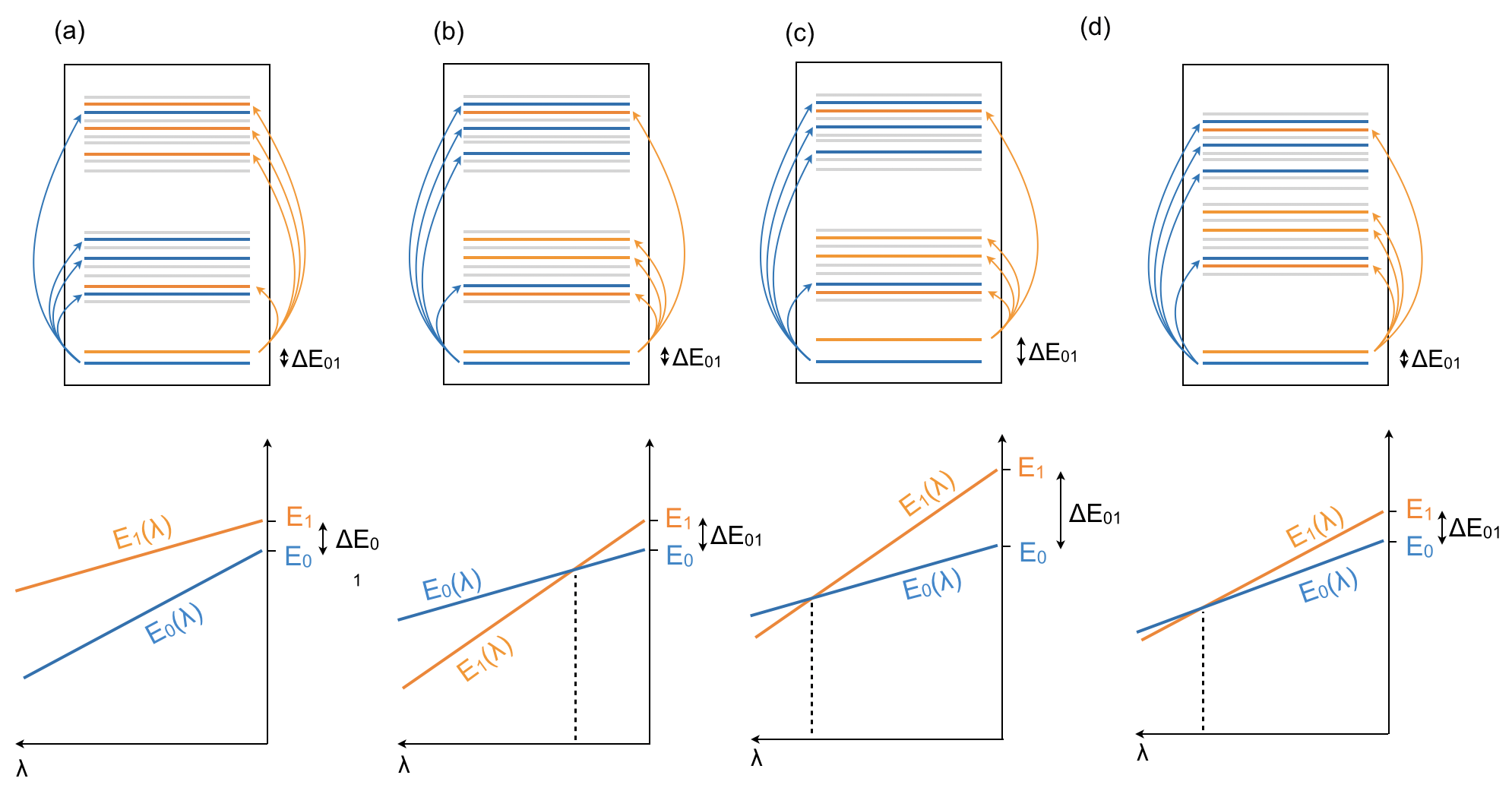}
    \caption{ Top: Cartoon problem energy spectra with the ground and first excited state, as well as their neighbourhoods, highlighted in blue and orange respectively. Bottom: Corresponding cartoons of the perturbed problem energies where the perturbation is the driver Hamiltonian and the perturbative parameter is $\lambda$. (a) $\ket{E_0}$ has a lower energy neighbourhood than $\ket{E_1}$ such that no perturbative crossing forms. (b)$\ket{E_1}$ has a lower energy neighbourhood than $\ket{E_0}$ such that a perturbative crossing does form. (c) Similar setting to (b) but with a larger $\Delta E_{01}$ such that the crossing happens at larger $\lambda$. (d) Similar setting to (b) but where the neighbourhoods are closer in energy, reducing the difference between the gradients of the perturbations, such that the crossing happens at larger $\lambda$. }
    \label{fig:cartoon-crossings}
\end{figure*}

Our problem instances are constructed to produce annealing spectra with perturbative crossings - a well understood bottleneck in QA that can arise when excited states of the problem Hamiltonian have an energy comparable to that of its ground-state \cite{Amin2009, Altshuler2010}. Their presence can be understood by treating the driver $H_d$ as a perturbation to the problem $H_p$, and looking at the energy corrections to the problem eigenstates. We briefly outline this technique before going on to describe our problem instances - more precise descriptions of how these ACs form can be found in  .

Writing the perturbed energies as $E_a(\lambda)$, we can say that a crossing occurs between two problem states $a$ and $b$ if $E_b(\lambda) < E_a(\lambda)$ ($b > a$) for some $\lambda$. If this $\lambda$ is small enough that perturbation theory remains valid, this indicates the formation of an AC towards the end of the instantaneous gap spectrum. Because the driver Hamiltonian couples problem states that are one spin-flip apart, the perturbation to the state $\ket{E_a}$ is dependent on the set of states that are within unit Hamming distance of $\ket{E_a}$ - we refer to this set as the \textit{neighbourhood} of $\ket{E_a}$. It can be shown that the lowest energy problem states will experience a reduction in energy as a result of the perturbation as long as the Hamming distance between any pair of these states is greater than 1. The rate of this reduction with respect to $\lambda$ for each state depends on the energy of the states in its neighbourhood, with lower energies resulting in a greater reduction. Thus a crossing can occur between the ground-state and some other problem state $\ket{E_a}$ that is close in energy if the neighbourhood around the state $\ket{E_a}$ contains lower energy states than that around the ground-state. The value of $\lambda$ for which the crossing occurs, and thus whether or not an AC forms, will depend on the difference in energy between the two unperturbed problem states, $\Delta E_{0a} = E_a - E_0$, and how different in energy the neighbourhoods around them are.

The top row of Figure \ref{fig:cartoon-crossings} shows cartoon problem energy spectra with $E_0$ and its neighbourhood highlighted in blue and $E_1$ and its neighbourhood in orange. (For clarity we include arrows showing the driver couplings from $\ket{E_0}$ and $\ket{E_1}$ to their respective neighbourhoods.) Corresponding cartoons illustrating the perturbed energies are shown below with the magnitude of the driver increasing from right to left.  Figure \ref{fig:cartoon-crossings}(a) depicts a setting where $\ket{E_0}$ has a lower energy neighbourhood than $\ket{E_1}$ such that it receives a greater negative perturbation and no crossing forms. Figure \ref{fig:cartoon-crossings}(b) depicts the case where the neighbourhoods are reversed such that $\ket{E_1}$ receives a greater negative perturbation and the perturbed energies cross. Figure \ref{fig:cartoon-crossings}(c) shows a similar scenario but with a greater $\Delta E_{01}$ such that a higher $\lambda$ is needed before the energies cross. Finally, in Figure \ref{fig:cartoon-crossings}(d) $\Delta E_{01}$ is the same as in Figure \ref{fig:cartoon-crossings}(a) but the difference between the energies of their neighbourhoods is decreased such that the difference between the gradients of the perturbed energies are smaller. This also increases the value of $\lambda$ for which the crossing occurs. 

In addition to changing the value of $\lambda$ for which the crossing occurs, and thus the value of $s$ at which we observe an AC in the annealing spectrum, $\Delta E_{01}$ and the neighbourhoods around $\ket{E_0}$ and $\ket{E_1}$ also affect the extent to which the perturbed state vectors $\ket{E_a(\lambda)}$ have evolved away from the problem state vectors $\ket{E_a}$ at the point of the crossing. In the case where the differences between $\ket{E_a(\lambda)}$ and $\ket{E_a}$ are minimal, the overlap between the instantaneous ground-state before and after the AC is very small. However if the perturbed problem states have become more mixed, then the overlap may be larger. The extent of this overlap affects the size of the gap minimum associated with the AC, with a larger overlap leading to a bigger gap \cite{Amin2008}.
\begin{figure}[]
    \centering
    \includegraphics[scale=0.33]{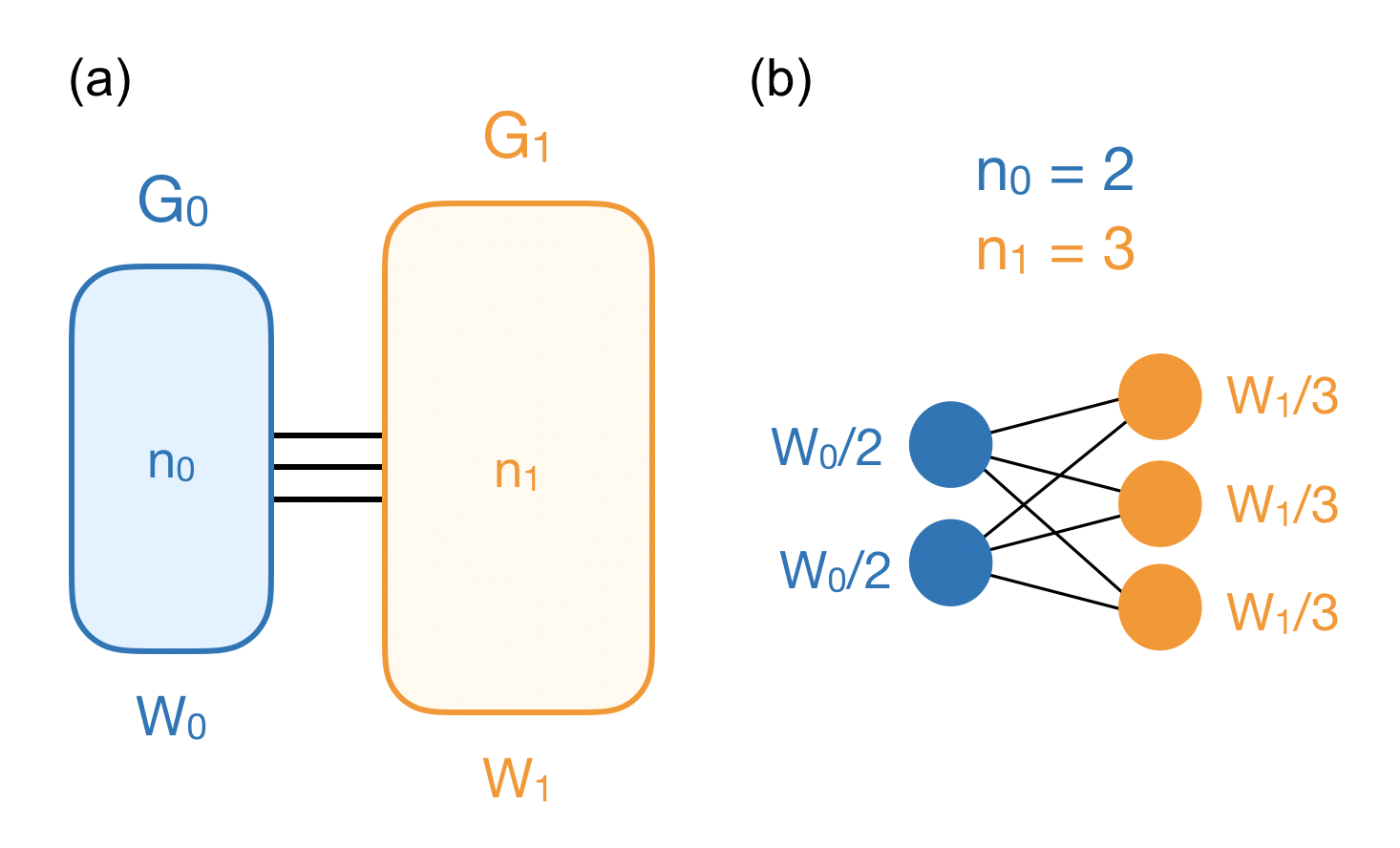}
    \caption{ An illustration of our graph structure, as described in Section \ref{sec:graph}, is shown in (a). An example with $n_0=2$ and $n_1=3$ is shown in (b). }
    \label{fig:problem-graphs}
\end{figure}

In order to produce an annealing spectrum with a perturbative crossing we thus require a problem setting where $E_0$ and $E_1$ are close in energy and where the neighbourhood around $\ket{E_1}$ contains lower energy states than that around $\ket{E_0}$. Additionally, if we wish to be able to adjust the gap size associated with the crossing, we need to be able to easily tune $\Delta E_{01}$ and the energy differences of the neighbourhoods. To achieve this, we utilise the MWIS problem, which takes as its input an undirected, weighted graph and aims to find the set of vertices with the largest weight for which no two vertices are connected by an edge. The problems we construct, as well as the ways we utilise the free parameters in its Ising Hamiltonian encoding, are inspired by the MWIS instances used in \cite{Choi2020}. We now describe how the MWIS problem Hamiltonian is constructed before going on to describe our specific problem setting.

In its Ising formulation, each vertex of the problem graph is represented by a spin. Each basis state then represents a set of vertices, with spin up denoting a vertex that is in the set and spin down denoting a vertex that is not.  For instance, the problem state $\ket{E_a} = \ket{\downarrow \uparrow \uparrow \downarrow \uparrow}$ corresponds to the set of vertices $\{2,3,5\}$. Note that for this encoding flipping a spin corresponds to either adding or removing a vertex from the set. The vertex weights are implemented with local Z-fields and the independent set condition by introducing an edge penalty. This penalty is achieved by adding an anti-ferromagnetic ZZ-coupling, with a strength $J_{zz}$, between any two qubits corresponding to vertices connected by an edge. Overall, the problem Hamiltonian is given by
\begin{equation}
\label{eq:Problem-Hamiltonian}
    H_p = \sum \limits_{i \in \{\textrm{\scriptsize vertices}\}} (c_i J_{zz} - 2w_i)\sigma^z_i + \sum \limits_{(i,j) \in \{\textrm{\scriptsize edges}\}} J_{zz} \sigma^z_i \sigma^z_j
\end{equation}
where $c_i$ is the number of edges connected to vertex $i$ and $w_i$ is the weight on vertex $i$. The fact that the edge penalty appears in the local field terms is to account for the fact that two neighbouring down spins should not be penalised. 

Assuming a high enough $J_{zz}$ is chosen, the energies of the problem states will form clusters based on how many edges are contained in the corresponding set of vertices. The higher the magnitude of $J_{zz}$, the greater the separation between these clusters. Within the clusters, the energies of the states will be ordered by the total weight of the corresponding set, with the highest weighted set having the lowest energy.

A diagram of the scalable MWIS problem used in this work is shown in Figure \ref{fig:problem-graphs}(a). It is a complete bipartite graph with $n_0$ vertices in sub-graph $G_0$ and $n_1$ vertices in sub-graph $G_1$. As a result, the graph has two maximally independent sets giving the MWIS problem on this graph two local optima. We allocate a weight to each sub-graph, $W_0$ and $W_1$, which we split equally between the vertices. We give each sub-graph a base weight of $W$ and then increase the weight on sub-graph $G_0$ by $\delta W$ such that $W_1 = W$ and $W_0 = W_1 + \delta W = W + \delta W$. So long as $\delta W$ is chosen to be small enough, this results in the set containing all the vertices in $G_0$ being the highest weighted independent set and the set containing all the vertices in $G_1$ being the independent set with the second highest weight. An example graph with $n_0=2$ and $n_1=3$ is shown in Figure \ref{fig:problem-graphs}(b). 

The relative sizes of the sub-graphs dictate the energy spectra of the neighbourhoods around $\ket{E_0}$ and $\ket{E_1}$. Recall that flipping one of the spins in $\ket{E_a}$ from down to up corresponds to adding a vertex to the corresponding set and that flipping a spin from up to down corresponds to removing a vertex. This means that the neighbourhood around $\ket{E_0}$ has $n_0$ neighbours corresponding to independent sets and $n_1$ neighbours corresponding to dependent sets - and vice versa for $\ket{E_1}$. Recalling further that the states corresponding to independent sets have lower energies than those corresponding to dependent sets, we can say that if $n_1 > n_0$ then $\ket{E_1}$ has more low energy neighbours than $\ket{E_0}$.

Given a particular problem graph, we can use $\delta W$ and $J_{zz}$ to tune $\Delta E_{01}$ and the differences between the energies of the neighbourhoods around $\ket{E_0}$ and $\ket{E_1}$. Roughly speaking, we use $\delta W$ to adjust $\Delta E_{01}$ and the edge penalty, $J_{zz}$, to change the separation between the clusters of problem energies associated with different numbers of edge violations. In practice however, the way in which we normalise $H_p$ to keep the energy scale consistent means that both parameters have some effect on all the energy gaps in the spectrum. We are interested in how the effect of an XX-catalyst changes depending on the strength of the AC and so we choose our values of $\delta W$ and $J_{zz}$ to adjust the extent to which the problem state vectors are perturbed before the crossing while keeping the value of $s$ for which the AC occurs for the $5$-vertex instance (on which we focus our discussion) unchanged. Further details on our Hamiltonian and how we select our parameters can be found in \ref{app:prop-parameter-tuning}.
\begin{figure*}[]
    \centering
    \includegraphics[scale=0.38]{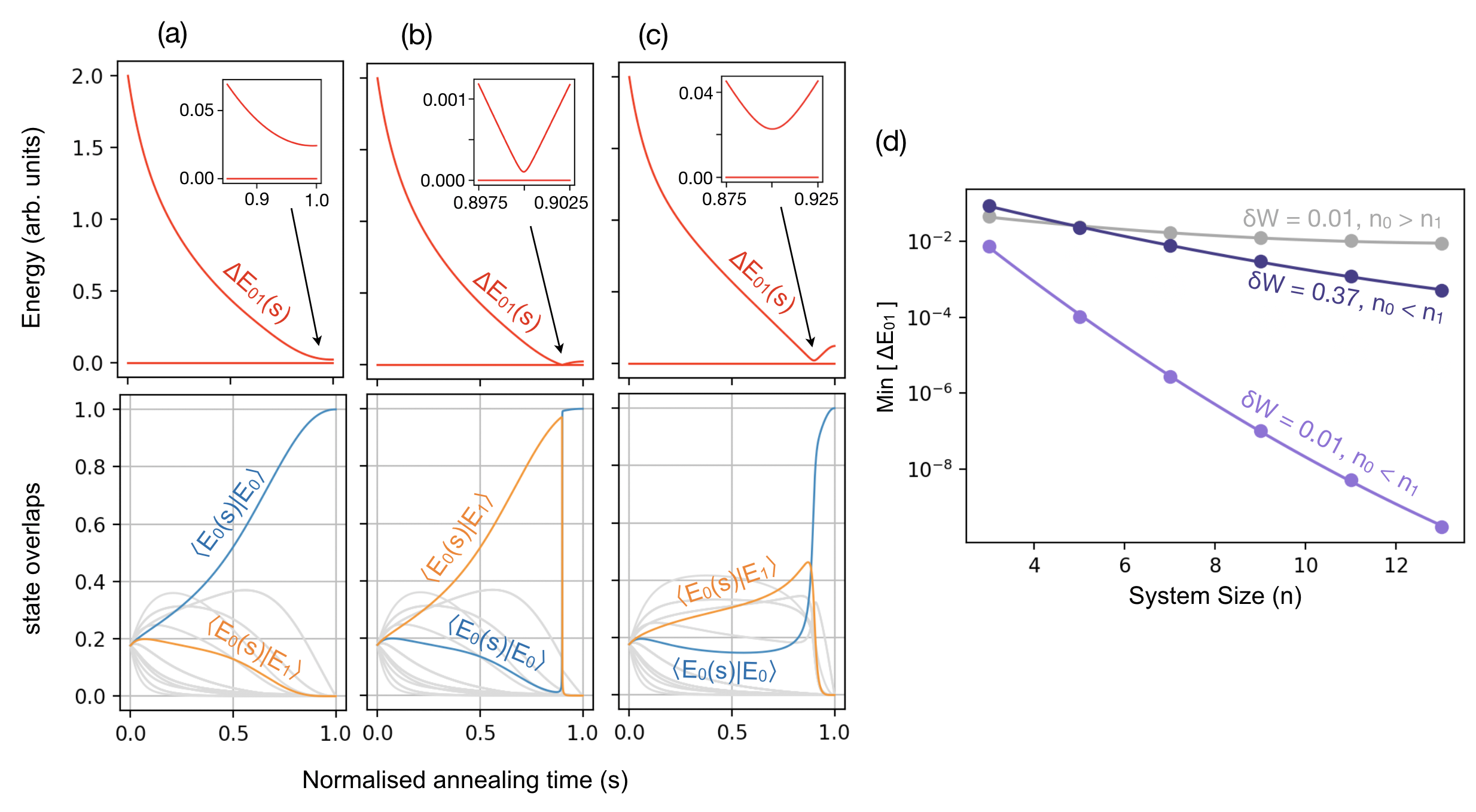}
    \caption{ (a-c) show numerical results for different anneals without the presence of a catalyst. Gap spectra are shown on the top and the evolution of the instantaneous ground-state vectors are shown on the bottom in terms of their overlaps with the problem states - the problem ground and first excited state overlaps are highlighted in blue and orange respectively. The problem parameters are: (a) $n_0=3$, $n_1=2$, $\delta W = 0.01$, $J_{zz}=5.33$. (b) $n _0=2$, $n_1=3$, $\delta W = 0.01$, $J_{zz}=5.33$ and (c) $n _0=2$, $n_1=3$, $\delta W = 0.37$, $J_{zz}=37.5$. (d) shows the minimum gap with increasing system size for three different parameter settings. The grey line corresponds to the parameter settings in (a) and the two purple lines correspond to the parameter settings in (b) and (c). For the cases where an AC is produced the sub-graph sizes are scaled as $n_0 = (n-1)/2$ and $n_1 = (n+1)/2$. For the case where we do not produce an AC the sub-graph sizes are reversed. Lines in (d) connecting the data points are a guide to the eye.}
    \label{fig:no-cat-results}
\end{figure*}
\begin{figure*}[]
    \centering
    \includegraphics[scale=0.35]{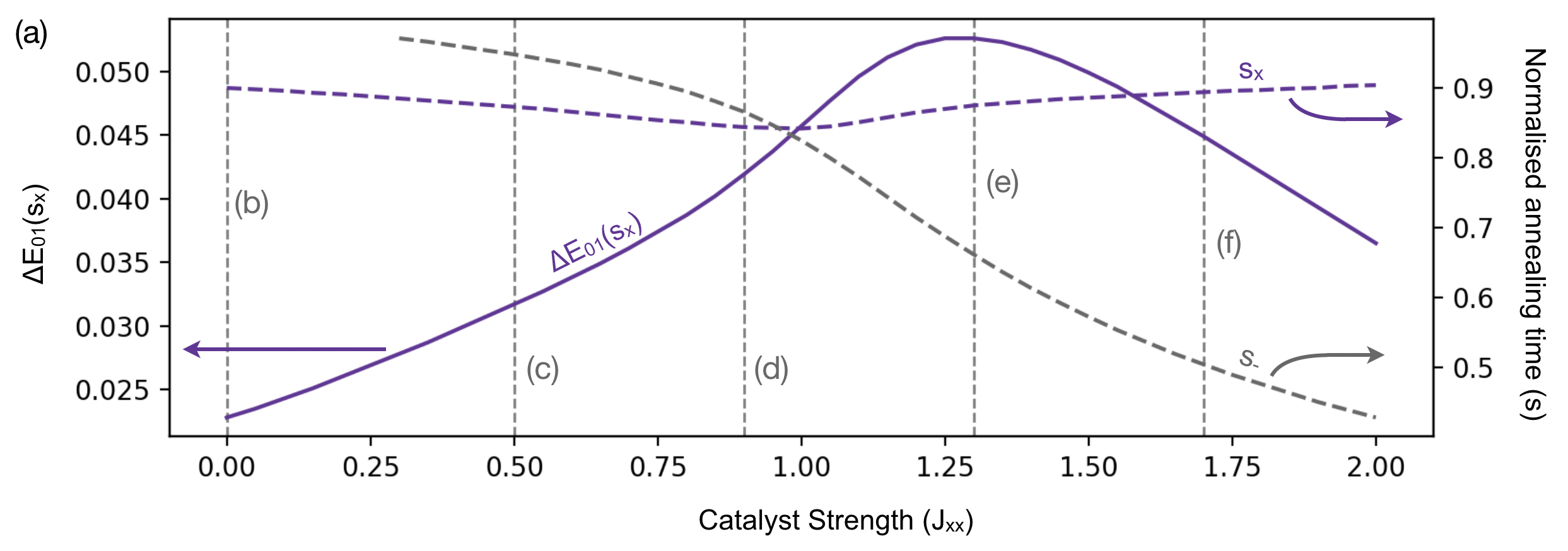}
    \includegraphics[scale=0.35]{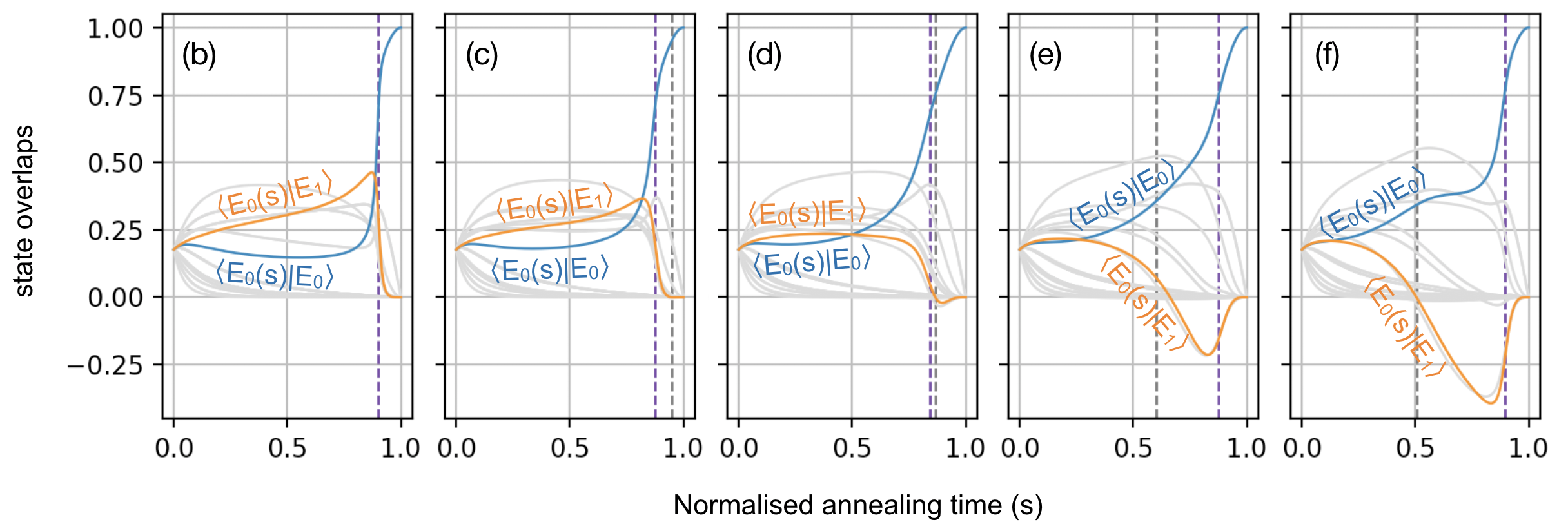}
    \caption{ Numerical results for a problem instance with $n_0=2$, $n_1=3$, $\delta W=0.37$ and $J_{zz}=37.5$ (i.e: the parameters associated with the WGS). A catalyst is applied between a single pair of vertices in sub-graph $G_1$. (a) shows the dependence on catalyst strength of the gap size at the AC, $\Delta E_{01}(s_\textrm{\scriptsize x})$ (solid purple), the location of the minimum gap, $s_\textrm{\scriptsize x}$ (dashed purple), and the value of $s$ for which $\braket{E_{0}(s)|E_1}$ becomes negative, $s_-$ (dashed grey). The evolution of the instantaneous ground-state for different catalyst strengths is shown in (b-f). These plots have $s_\textrm{\scriptsize x}$ and $s_-$ marked with purple and grey dashed lines respectively. The catalyst strengths for which we show the evolution are marked on (a) with vertical grey dashed lines. }
    \label{fig:cat-results-1}
\end{figure*}

\subsection{ Spectral Properties \label{sec:parameters} }

We now present numerical results corresponding to different parameter settings to confirm that we produce the desired AC and that we are able to alter its properties in the way described above. The results that we plot are the energy gaps between the instantaneous ground and first excited state, $\Delta E_{01}(s)$, as well as the evolution of the instantaneous ground-state, $\ket{E_0(s)}$. The evolution of $\ket{E_0(s)}$ we consider in terms of its overlaps with the problem states - in particular with $\ket{E_0}$ and $\ket{E_1}$ since it is between these two states that we expect a crossing. Our results are obtained by numerical diagonalisation of $H(s)$ at different values of $s$ with no catalyst present. 

We plot on the top row of figures \ref{fig:no-cat-results}(a)-(c) the energy difference $\Delta E_{01}$ as a function of $s$. On the bottom row of figures \ref{fig:no-cat-results}(a)-(c) we show the corresponding evolution of $\bra{E_0(s)} E_{0,1} \rangle$. In Figure \ref{fig:no-cat-results}(a) we show the case where $n_0 > n_1$ for a given set of problem parameters, which results in the absence of an AC since $\ket{E_0}$ has a lower energy neighbourhood than $\ket{E_1}$. Figure \ref{fig:no-cat-results}(b) shows the case where $n_1 > n_0$ but with the other problem parameters ($\delta W$ and $J_{zz}$) kept the same as in (a). This now results in an AC occurring at $s=0.9$ since $\ket{E_1}$ now has a lower energy neighbourhood than $\ket{E_0}$. In both (a) and (b), the MWIS problem parameters are $\delta W = 0.01$ and $J_{zz} = 5.33$. Figure \ref{fig:no-cat-results}(c) illustrates the case where $n_1 > n_0$, but the MWIS problem parameters have been increased to $\delta W = 0.37$ and $J_{zz} = 37.5$. This results in an increase of $\Delta E_{01}$ and enhances the differences between the energies of the neighbourhoods of $\ket{E_0}$ and $\ket{E_1}$. As in Figure \ref{fig:no-cat-results}(c), we observe a gap minimum at $s=0.9$ indicative of an AC.

In the bottom plots of Figures \ref{fig:no-cat-results}(b) and (c) we see that the instantaneous ground-state evolves towards $\ket{E_1}$ before there is a sharp exchange at $s=0.9$, after which $\ket{E_0(s)}$ becomes dominated by $\ket{E_0}$. This exchange, coinciding with the gap minimum, is what we expect to see at a perturbative crossing \cite{Choi2010}. This is in contrast to the bottom plot in Figure \ref{fig:no-cat-results}(a) where the instantaneous ground-state is seen to evolve smoothly towards $\ket{E_0}$. Finally, in Figure \ref{fig:no-cat-results}(d), we show the scaling of the gap minimum with system size. For the two cases where an AC is produced we scale the sub-graph sizes as $n_0 = (n-1)/2$ and $n_1 = (n+1)/2$ and for the case where we do not produce an AC these sizes are reversed. We use, for each system size, the same values of $\delta W$ and $J_{zz}$ as for our $5$ vertex examples. We see that the minimum gap size does indeed appear to close exponentially in the cases where $n_0 < n_1$ - further indicating the presence of an AC.

Figures \ref{fig:no-cat-results}(a)-(d) show that the AC can be manipulated in the desired way by changing our problem parameters. First, we are able to control whether or not an AC occurs in the spectrum by changing the relative sizes of $n_0$ and $n_1$ as intended. Secondly, we are able to adjust the extent to which the instantaneous states have evolved away from the problem states at the point of the crossing by changing $\delta W$ and $J_{zz}$. This alters the scaling of the minimum gap against system size. We observe that as $\delta W$ and $J_{zz}$ are increased for $n_1 > n_0$, the gap does still appear to close exponentially, but the scaling has been `weakened', i.e. the scaling exponent is reduced. Therefore, in the following, we will refer to our two problem parameter settings (which produce an AC) as the weak gap scaling (WGS) and strong gap scaling (SGS) cases.

\section{ XX-Catalysts \label{sec:results} }

With our problem setting established, we now turn to the effects of introducing an XX-catalyst into the annealing Hamiltonian. As stated in Section \ref{sec:hamiltonian}, the catalysts used in this work contain a single XX-coupling which is introduced into $H(s)$ with the opposite sign to the driver and a strength of $|J_{xx}|$ (equation \ref{eq:Hc}). This results in a non-stoquastic $H(s)$ for $s \neq 0,1$ and as such it is possible for components of the instantaneous GS vector to become negative during the anneal. That the relative signs of these vector components can be key to understanding the structure of quantum annealing spectra was highlighted in \cite{Choi2021} - and we will see that they also play a crucial role in our results.

The XX-coupling is applied between a single pair of vertices in $G_1$. (Permutation symmetry within the sub-graphs means that there is no need to specify which two vertices are selected.) The result of this XX term is to couple $\ket{E_1}$ to a state corresponding to an independent set and $\ket{E_0}$ to a state corresponding to a dependent set - that is, the catalyst couples $\ket{E_1}$ to a state with significantly lower energy than that to which it couples $\ket{E_0}$. 

This approach can intuitively be thought of as a counter-approach to giving $\ket{E_0}$ additional low energy neighbours with a \textit{stoquastic} catalyst - i.e one that enters with the same sign as the driver. This stoquastic approach can be understood through the same arguments as those made in Section \ref{sec:graph} regarding the formation of perturbative crossings and was shown in \cite{Choi2020} to enhance the minimum gap size. There is some numerical evidence that taking the reverse approach with a non-stoquastic catalyst (i.e one that has been introduced with the opposite sign to the driver) is able to weaken or remove a perturbative crossing \cite{Choi2021,Albash2019} - although in these examples the catalysts are specifically coupling together local optima and so are not a perfect analogy to the work in \cite{Choi2020}). \cite{Choi2021} also introduces a subtly different perturbative argument to explain why such non-stoquastic catalysts are able to enhance the gap size at an AC. We apply the same perturbative approach to our setting in \ref{app:pert_argument} in order to further support our choice of XX-coupling.

\subsection{ Weak Gap Scaling Case \label{sec:enhanement} }

\begin{figure*}[]
    \centering
    \includegraphics[scale=0.32]{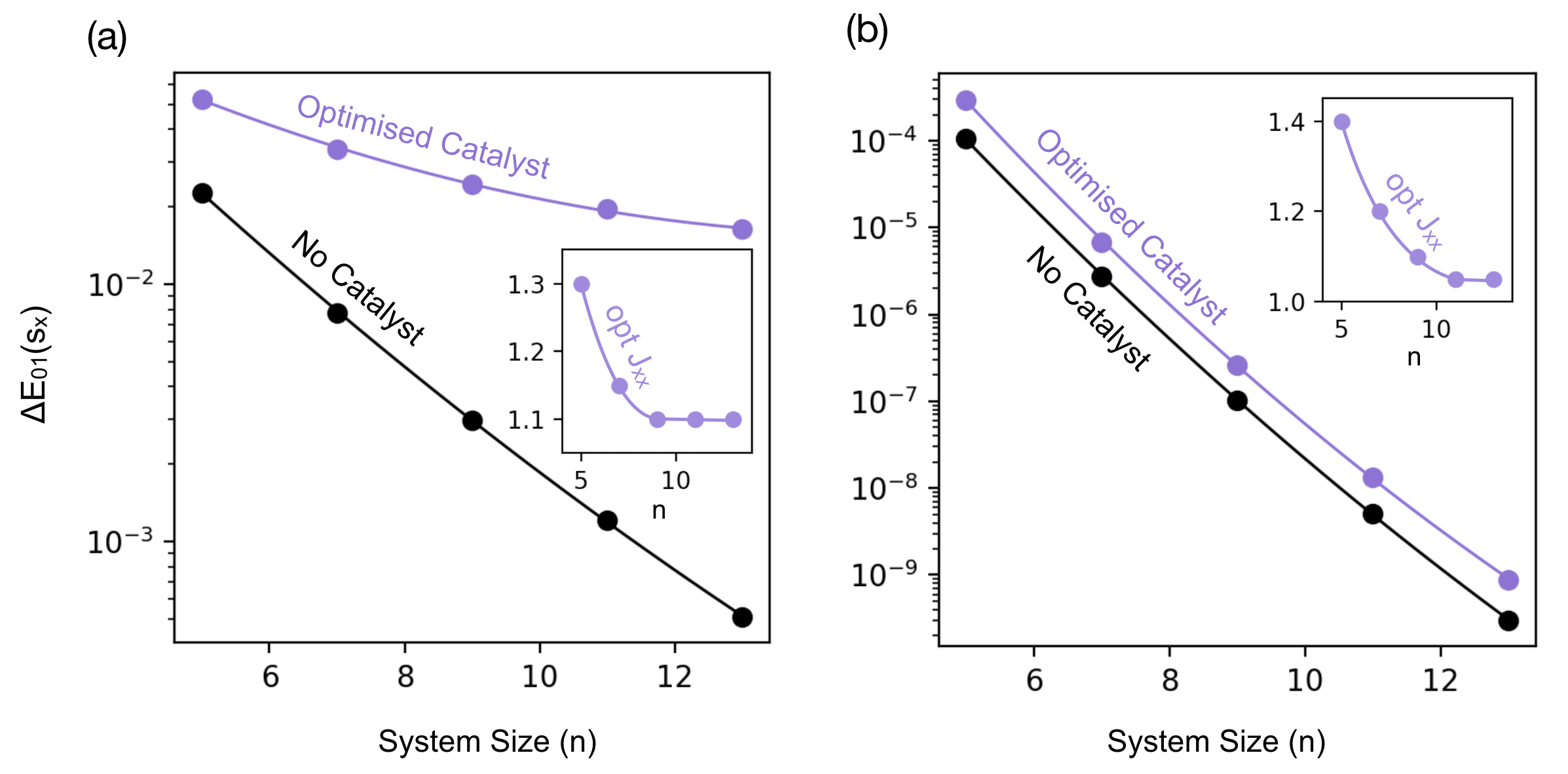}
    \caption{ Results for the size of the gap minimum at the AC with increasing system size. The sub-graph sizes are scaled as $n_0 = (n-1)/2$ and $n_1 = (n+1)/2$. (a) shows results for the WGS setting ($\delta W=0.37$ and $J_{zz}=37.5$) and (b) shows results for the SGS setting ($\delta W=0.01$ and $J_{zz}=5.33$). In each case we show the gap size without a catalyst in black and with $J_{xx}$ chosen to maximise the gap size in purple. The values of $J_{xx}$ that maximise the gap for each $n$ are plotted in the insets. Lines connecting the data points are a guide to the eye.}
    \label{fig:cat-gap-scaling}
\end{figure*}

In Figure \ref{fig:cat-results-1} we present numerical results of varying the XX catalyst strength $J_{xx}$ on the annealing problem described in Section \ref{sec:parameters}, with $n_0=2$, $n_1=3$, $\delta W = 0.37$ and $J_{zz}=37.5$. Recall that, in the catalyst-free case, these parameter choices result in the comparatively weak AC shown in Fig \ref{fig:no-cat-results}(c). Figure \ref{fig:cat-results-1}(a) shows the dependence of the minimum gap, $\Delta E_{01}(s_\textrm{\scriptsize x})$, as a function of the catalyst strength $J_{xx}$. Also shown is the variation of the minimum gap location, $s_\textrm{\scriptsize x}$, and the location where $\bra{E_0(s)} E_{1} \rangle$ becomes negative, $s_-$. Figures \ref{fig:cat-results-1}(b)-(f) illustrate the behavior of the coefficients $\bra{E_0(s)} E_{0,1} \rangle$ at selected values of $J_{xx}$.  It can be seen that $\braket{E_0(s)| E_{1}}$ can become negative at sufficiently large values of $J_{xx}$. We observe that the introduction of the catalyst results in a clear maximum value of $\Delta E_{01}(s_\textrm{\scriptsize x})$ at $J_{xx} = 1.30$. The existence of an optimal $J_{xx}$, rather than a monotonic improvement, was also observed in other work \cite{Albash2019,Choi2021}.

We obtain similar results when increasing the system size. Figure \ref{fig:cat-gap-scaling}(a) compares the gap scaling without the catalyst (black) and the gap scaling when we use a catalyst with $J_{xx}$ optimised for each system size (purple) - the inset shows the optimal values of $J_{xx}$ for each system size. We see that by using the optimal values we are able to greatly reduce the severity of the gap scaling with the optimised scaling appearing to be sub-exponential - although for these system sizes it is difficult to discern what the true scaling behaviour is. We also note that the optimal $J_{xx}$ values are seen to quickly tend to a constant value as the system size is increased despite the fact that we are using a single coupling in our catalyst. This point will be discussed further in Section \ref{sec:discussion}.

Turning our attention to the evolution of the problem state overlaps $\bra{E_0(s)} E_{0,1} \rangle$ shown in Figures \ref{fig:cat-results-1}(b-f), we observe that changes in $\bra{E_0(s)} E_{0,1} \rangle$ are overall less significant when $\Delta E_{01}(s_\textrm{\scriptsize x})$ is largest. The relation between the rate of change of the instantaneous ground-state and the gap separating it from $\ket{E_1(s)}$ is well understood and expressions explicitly linking them can be found in \cite{Braida2021}. 

\subsection{ Strong Gap Scaling Case \label{sec:closing_gaps} }

\begin{figure*}[]
    \centering
    \includegraphics[scale=0.39]{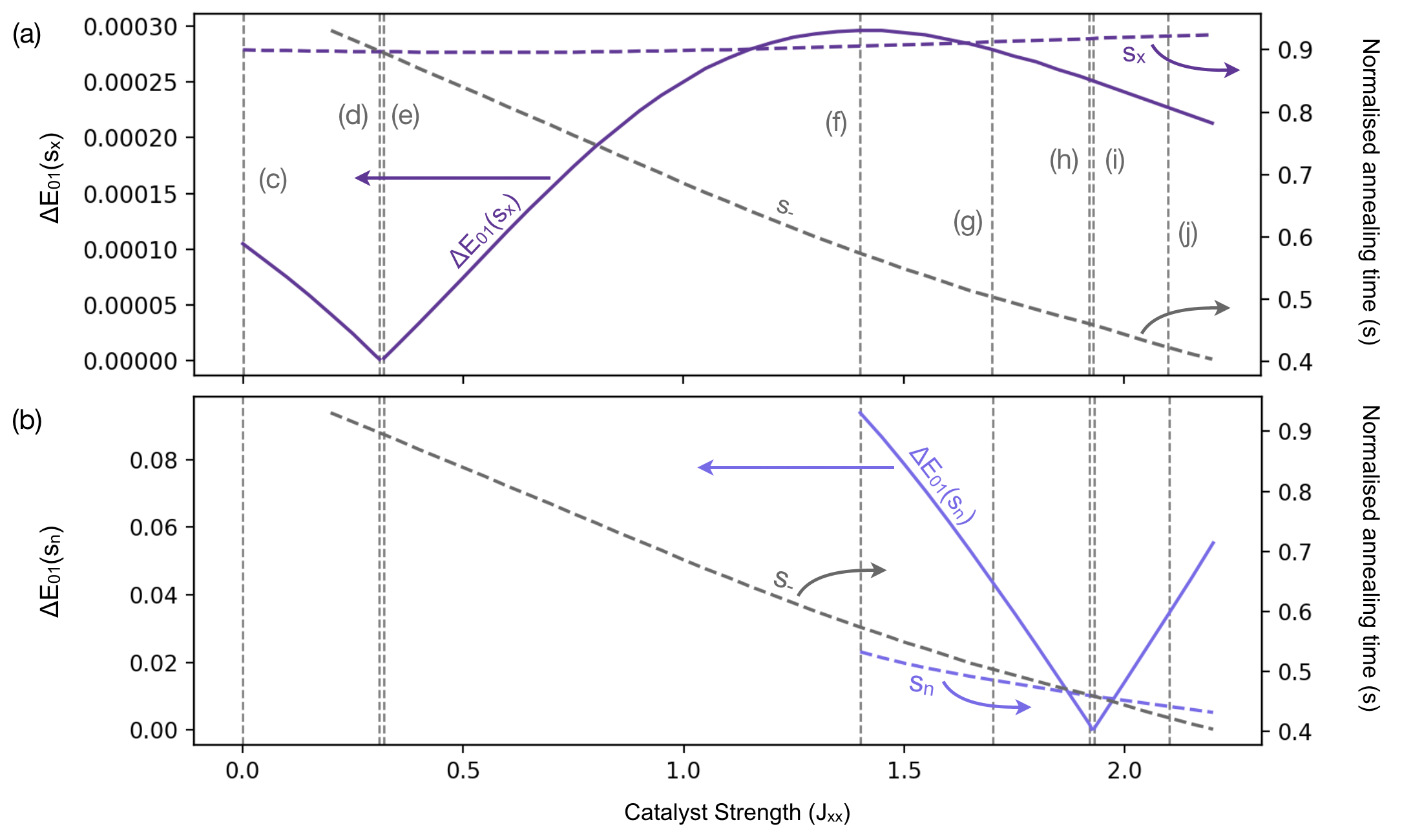}
    \includegraphics[scale=0.39]{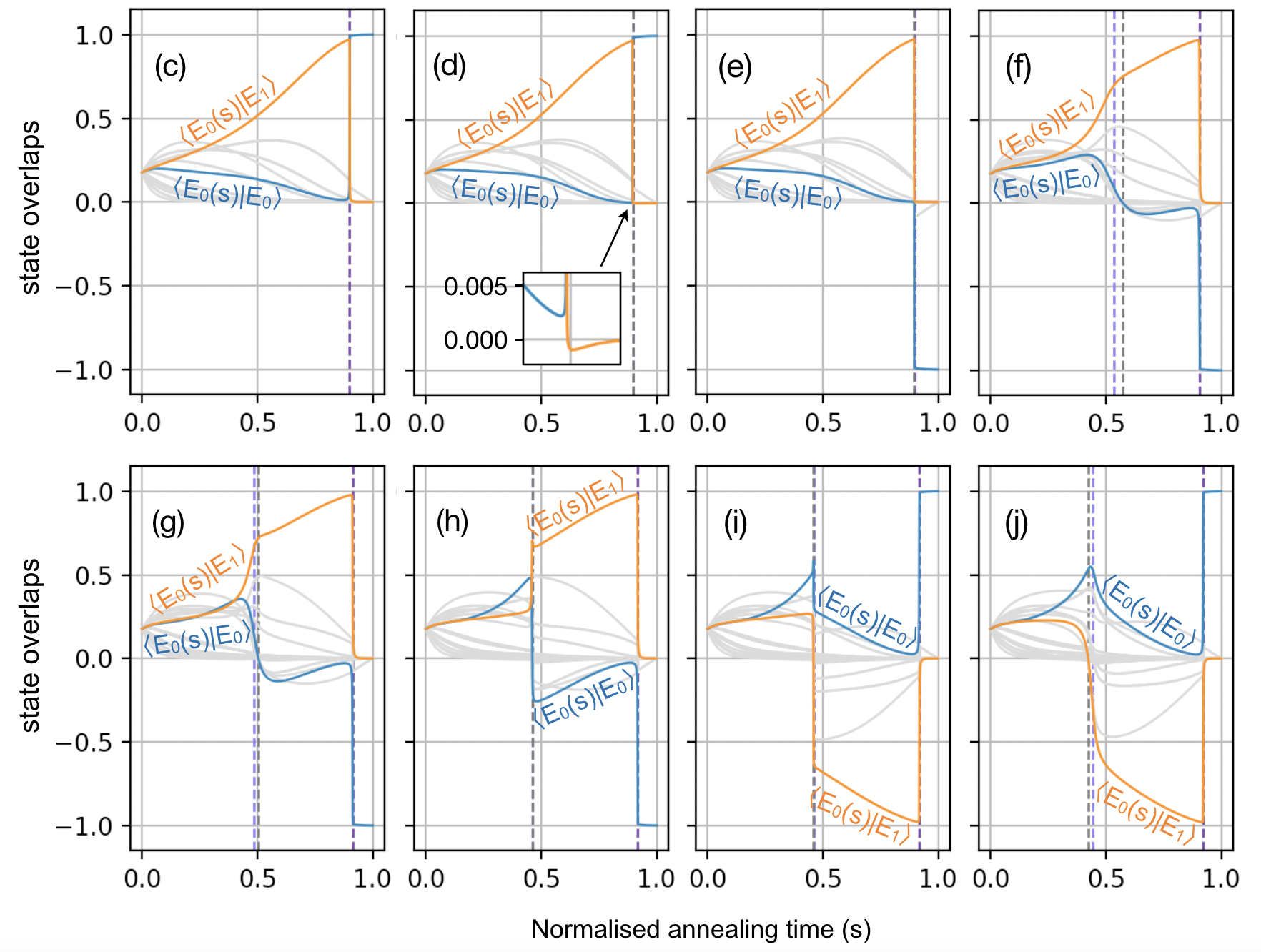}
    \caption{ Numerical results for a problem instance with $n_0=2$, $n_1=3$, $\delta W=0.01$ and $J_{zz}=5.33$ (i.e: the parameters associated with the SGS). A catalyst is applied between a single pair of vertices in sub-graph $G_1$. (a) shows, for increasing catalyst strength, results for the gap size at the AC, $\Delta E_{01}(s_\textrm{\scriptsize x})$ (solid purple), the location of the minimum gap, $s_\textrm{\scriptsize x}$ (dashed purple), and the value of $s$ for which either $\braket{E_{0}(s)|E_0}$ or $\braket{E_{0}(s)|E_1}$ becomes negative, $s_-$ (dashed grey). The lower plot shows, for the same $J_{xx}$ values, the gap size of the additional minimum gap that forms, $\Delta E_{01}(s_\textrm{\scriptsize n})$ (solid lighter purple), the location of this additional minimum, $s_\textrm{\scriptsize n}$ (dashed lighter purple) and $s_-$ (dashed grey). The evolution of the instantaneous ground-state for different catalyst strengths is shown in (c-j).  These plots have $s_\textrm{\scriptsize x}$, $s_\textrm{\scriptsize n}$ and $s_-$ marked with the deeper purple, the lighter purple and grey respectively. The catalyst strengths for which we show the evolution are marked on (a) and (b) with vertical grey dashed lines. }
    \label{fig:cat-results-2}
\end{figure*}
\begin{figure*}[]
    \centering
    \includegraphics[scale=0.35]{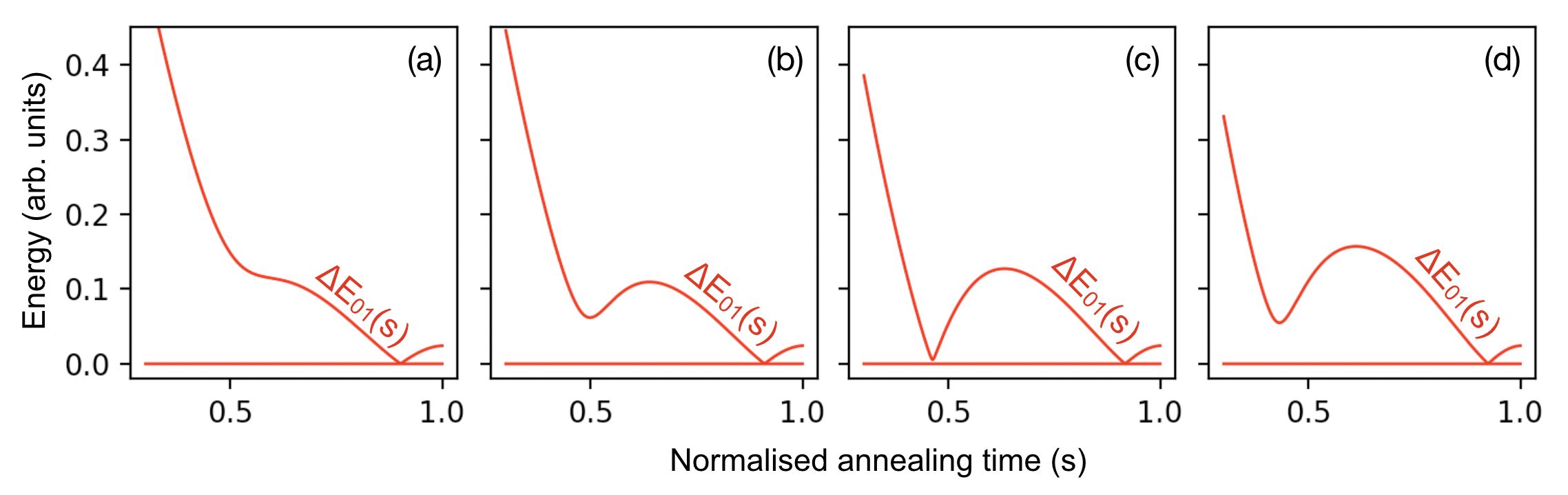}
    \caption{ Gap spectra for four different catalyst strengths showing the formation of the additional minimum gap between the instantaneous ground and first excited state in the SGS setting. The problem parameters are $n_0=2$, $n_1=3$, $\delta W=0.01$ and $J_{zz}=5.33$ and the catalyst is applied between two vertices in sub-graph $G_1$. The catalyst strengths are $J_{xx} = $ (a) $1.3$ (b) $1.6$ (c) $1.9$ (d) $2.2$ }
    \label{fig:new-gap-min}
\end{figure*}
We now examine the effect of this same catalyst in the SGS setting, i.e. for the MWIS problems with $\delta W=0.01$ and $J_{zz}=5.33$. Figure \ref{fig:cat-gap-scaling}(b) shows the scaling of the minimum gap in the SGS case as a function of system size with and without the catalyst in purple and black respectively. As in the WGS case, the value of $J_{xx}$ used for each $n$ is chosen to maximise the gap enhancement and is shown in the inset. An important difference when compared to the WGS case is that while we do still observe some gap enhancement at the AC, the catalyst does not significantly alter the scaling with system size. This suggests that, unlike in the WGS case, the use of such an XX catalyst in this situation is of little use for alleviating the exponential drop-off in fidelity due to the exponentially closing gap.

Numerical results for the size of the gap minimum at the AC for the $5$-vertex instance at different catalyst strengths are presented in Figure \ref{fig:cat-results-2}(a). The location of this gap minimum is denoted as $s_\textrm{\scriptsize x}$ as before. We see that, as for the weaker AC, the gap size does reach a maximum for an optimum catalyst strength. However, we now observe a closing of the gap for $J_{xx} \approx 0.3$. In addition to this we also observe the formation of a new, second local gap minimum earlier in the anneal, at a location we denote as $s_\textrm{\scriptsize n}$. The dependence of the size of this new local gap minimum on $J_{xx}$ is shown in Figure \ref{fig:cat-results-2}(b). To illustrate the manifestation of this second local gap minimum during an anneal, we plot the evolution of $\Delta E_{01}(s)$ for different catalyst strengths in Figure \ref{fig:new-gap-min}, which clearly shows the occurrence of a double local energy minima in the annealing spectrum. As with the gap minimum at $s_\textrm{\scriptsize x}$, we appear to be able to bring this gap at $s_\textrm{\scriptsize n}$ arbitrarily close to zero, albeit at a different catalyst amplitude. Note that the plots in Figures \ref{fig:cat-results-2}(a)-(b) correspond to exactly the same setting and that our separating the results for the two gap minima into different plots is only for readability of the data.

As in Figure \ref{fig:cat-results-1}(a), Figure \ref{fig:cat-results-2}(a) also shows the location of the gap minimum, $s_\textrm{\scriptsize x}$ with a dashed purple line, and the point at which the sign change in ground-state vector components occurs, $s_-$, with a dashed grey line. As $J_{xx}$ increases, the value of $s_-$ decreases which can be understood by the fact that this change in signs is associated with the non-stoquasticity introduced by the catalyst. This argument can be made more explicit using ideas introduced in \cite{Choi2021}. We see that the closing gap is observed for the $J_{xx}$ value at which $s_- = s_\textrm{\scriptsize x}$. Figure \ref{fig:cat-results-2}(b) shows the same data but for the additional gap minimum that forms in the strong AC setting. As with the previously discussed gap minimum, this new gap minimum approaches zero for the $J_{xx}$ value at which $s_- = s_n$. Associating the closing gaps with values of $J_{xx}$ at which the minimum gap locations coincide with $s_-$ is however not sufficient to account for the differing behaviours in the WGS and SGS settings. In Figure \ref{fig:cat-results-1}(a) from the preceding section, we see that there is also a value of $J_{xx}$ for which $s_- = s_\textrm{\scriptsize x}$ however no closing of the gap at the AC is observed.

To better understand the differences in the behavior between these two regimes we compare how the evolution of the ground-state vector changes with $J_{xx}$. Similarly to the WGS setting, we show the behavior of the overlaps $\bra{E_0(s)} E_{0,1} \rangle$ at specific values of $J_{xx}$ in Figures \ref{fig:cat-results-2}(c)-(j). In both the WGS (Figures \ref{fig:cat-results-1}(b)-(f)) and SGS (Figures \ref{fig:cat-results-2}(c)-(j)) cases, we clearly see that the components corresponding to the problem ground and first excited states take opposite signs at some $s$ given sufficiently large $J_{xx}$. We note however that, in contrast to the WGS case where it is always $\braket{E_{0}(s)|E_1}$ that becomes negative, whether $\braket{E_{0}(s)|E_0}$ or $\braket{E_{0}(s)|E_1}$ takes a negative value in the SGS case depends on the value of $J_{xx}$. In each plot in Figures \ref{fig:cat-results-2}(c)-(j), $s_\textrm{\scriptsize x}$, $s_n$ and $s_-$ are marked in dark purple, light purple and grey respectively and the $J_{xx}$ values to which these plots correspond are marked in Figure \ref{fig:cat-results-2}(a) with vertical dashed grey lines. Looking at the plots corresponding to $J_{xx}=0$ and $0.31$, Figures \ref{fig:cat-results-2}(c) and (d), we see that as we increase $J_{xx}$ towards the value at which $E_{01}(s_\textrm{\scriptsize x})$ vanishes, there is a sharpening of the change in the vector components around $s_\textrm{\scriptsize x}$. Then, as $s_-$ passes $s_\textrm{\scriptsize x}$, which happens between Figures \ref{fig:cat-results-2}(d) and (e), there appears to be a discontinuous change in which overlap crosses zero. After this point, the change in $\ket{E_0(s)}$ around $s_\textrm{\scriptsize x}$ begins to soften again. Looking at the Figures \ref{fig:cat-results-2}(f)-(j), we see the same sharpening (up to  $J_{xx}=1.92$ in Figure \ref{fig:cat-results-2}(h)) and then weakening of the rate of change in $\ket{E_0(s)}$ around the location of the new gap minimum, $s_n$. We also observe another shift in which vector component crosses zero as $s_-$ passes $s_n$; this occurs between Figures \ref{fig:cat-results-2}(h)-(i). 

This behaviour is in stark contrast to the WGS case where no distinct changes are observed around the $J_{xx}$ value where $s_- = s_\textrm{\scriptsize x}$. It is unclear why exactly this should be the case. However we can comment on some differences we observe between the two settings. In the WGS case, the AC has already become significantly smoothed out for the value of $J_{xx}$ at which $s_- = s_\textrm{\scriptsize x}$. This is in contrast to the SGS setting where the magnitudes of $|\braket{E_0(s)|E_0}|$ and $|\braket{E_0(s)|E_1}|$ that are exchanged at the AC are largely unchanged for the $J_{xx}$ value at which $s_- = s_\textrm{\scriptsize x}$. This is at least partially explained by the fact that if the initial AC is weaker, we can expect that a smaller change to the Hamiltonian is required to lift it. We also note that $s_-$ does not appear until higher values of $J_{xx}$ in the WGS setting and that the catalyst strength for which $s_-$ passes $s_\textrm{\scriptsize x}$ is over three times as high as it is for the SGS case. It may therefore be possible to lift the AC to a greater extent before reaching the value of $J_{xx}$ for which $s_- = s_{x}$.

\subsection{ Intermediate Regime \label{sec:intermediate}}

\begin{figure}[]
    \centering
    \includegraphics[scale=0.25]{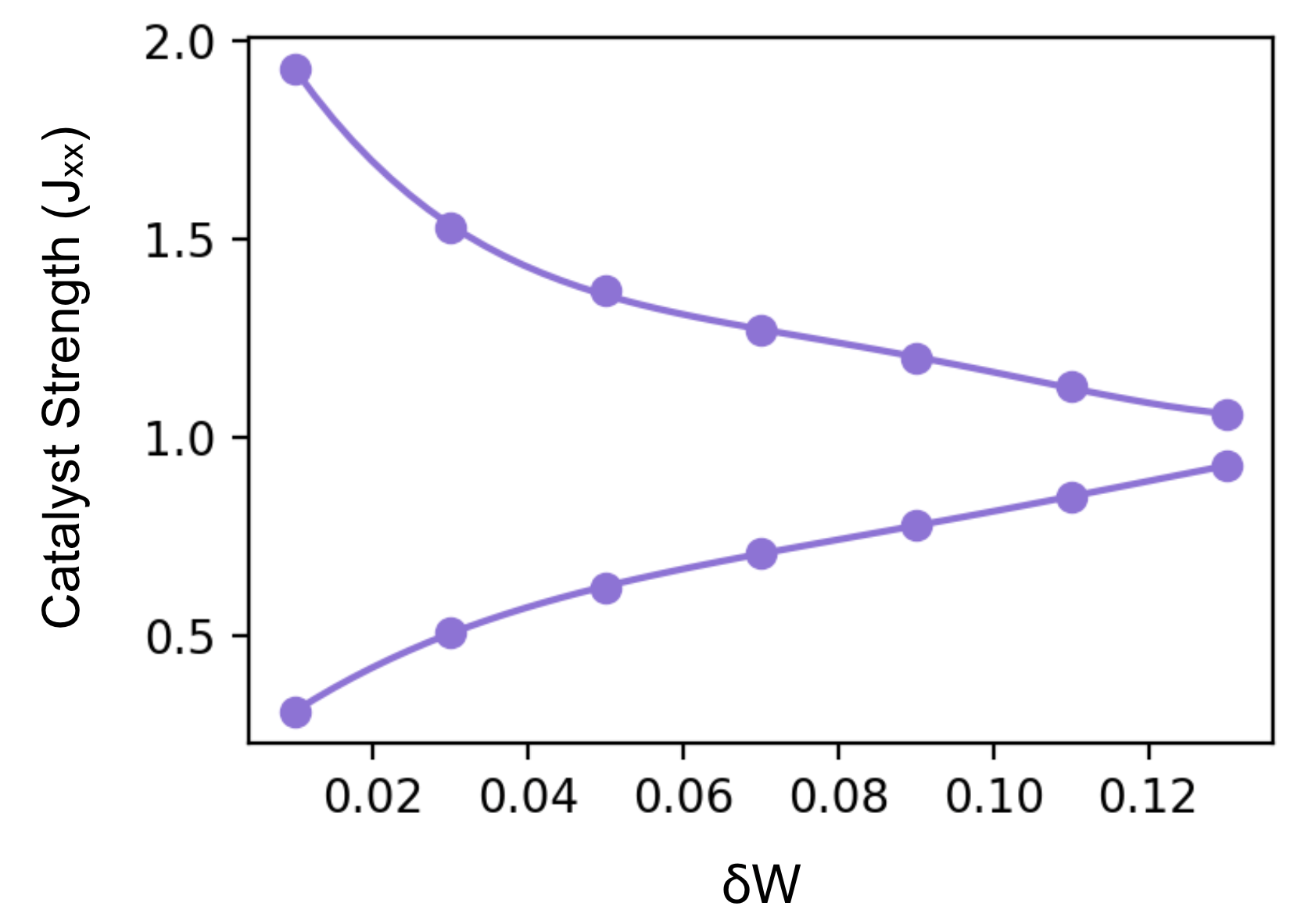}
    \caption{ $J_{xx}$ at which the two gaps go to zero for different parameter settings. The x-axis shows the $\delta W$ values for each parameter setting. The corresponding $J_{zz}$ value for each case is chosen such that $s_\textrm{\scriptsize x} = 0.9$. Lines connecting the data points are a guide to the eye. }
    \label{fig:Jxx-dW}
\end{figure}

We now present data examining intermediate parameter settings to further understand the changing behaviour as we move from the SGS to the WGS case. In Figure \ref{fig:Jxx-dW} we plot the $J_{xx}$ values associated with the two closing gaps for the 5-vertex instance with seven different parameter settings. The x-axis gives the $\delta W$ used for each parameter setting. Note however that the edge penalty, $J_{zz}$, is also changing between each data point, with its value in each case chosen such that $s_\textrm{\scriptsize x}=0.9$ without the presence of a catalyst. We observe that the two $J_{xx}$ values approach each other as we adjust the parameters to weaken the AC present in the original annealing spectrum until both values disappear at $\delta W \approx 0.13$. Given that the closing gaps are associated with a change in which vector component changes sign, one interpretation of these results could be that the lack of closing gaps in the weak AC setting is due to the vanishing of the $J_{xx}$ range for which the vector component corresponding to the problem ground-state crosses zero.

\section{ Discussion \label{sec:discussion} }

\begin{figure}[]
    \centering
    \includegraphics[scale=0.34]{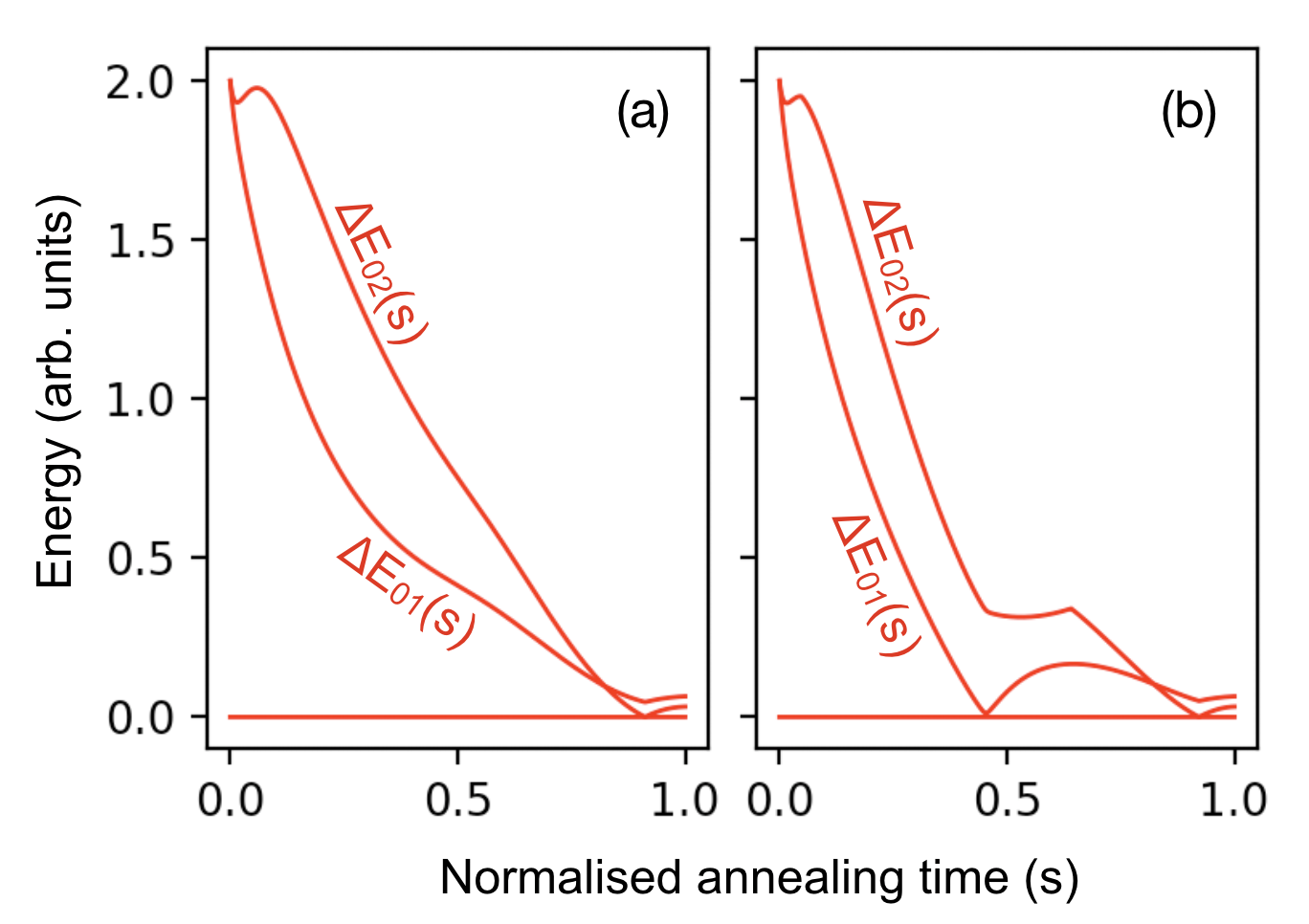}
    \caption{ Gap spectra for an anneal corresponding to a tri-partite MWIS problem with two local optima in addition to the global optimum. The results with and without a catalyst are shown in (b) and (a) respectively. Details on the construction of this problem instance are given in \ref{app:mult-opt}. }
    \label{fig:mult-local-opt}
\end{figure}
\begin{figure}[]
    \centering
    \includegraphics[scale=0.32]{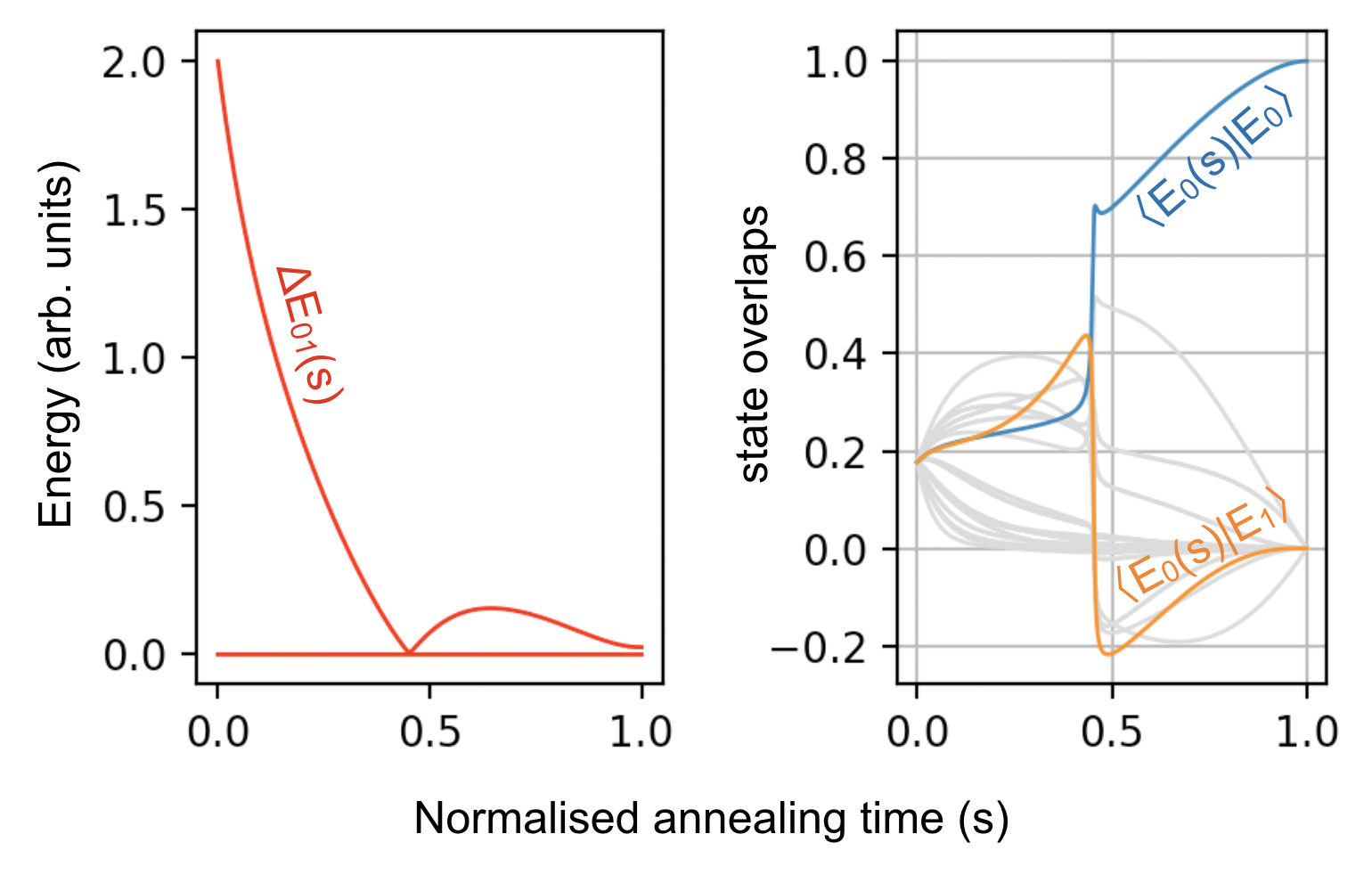}
    \caption{ Gap spectrum and ground-state evolution for a problem instance with $n_0=3$, $n_1=2$, $\delta W=0.01$ and $J_{zz} = 5.33$ and a catalyst containing one coupling within $G_0$. Results for this problem instance without a catalyst can be found in Figure \ref{fig:no-cat-results}(a). }
    \label{fig:no-AC-case}
\end{figure}

An important effect we have observed is that, while the targeted catalyst was able to suppress the gap scaling in the WGS case, it did not result in the intended suppression to the gap scaling in the SGS setting. The resulting spectrum in the latter case, however, appears amenable to diabatic quantum annealing (DQA)\cite{Crosson2020, Fry-Bouriaux2021, Choi2021}. Rather than running the algorithm slowly enough that the system remains in the ground-state, the system could be allowed to transition into the first excited state at the first small gap (the new gap minimum produced by the catalyst) and then back into the ground-state at the second (the gap minimum corresponding to the AC present in the original annealing spectrum). Simulations of the dynamics of the $5$-qubit problem in a closed-system setting suggest that this is indeed possible and results in a significant reduction of the annealing time required to reach the problem ground-state with high fidelity. These results are presented in \ref{app:DQA}.

So far we have only considered the effect of the catalyst on annealing spectra corresponding to the very simple MWIS graph shown in Figure \ref{fig:problem-graphs}. This was so we could examine the differing effects from the catalyst in the most straightforward setting. However, note that we observe these closing gaps for more general MWIS instances with additional local optima such that the problem Hamiltonian has more excited states with energy comparable to that of the ground-state. These instances can be constructed in much the same way as the problem graph shown in Figure \ref{fig:problem-graphs}(a), but with additional sub-graphs such that we have a complete k-partite graph. The catalysts were applied to instances with up to four sub-graphs and in each case the coupling was chosen to be within the sub-graph corresponding to the excited problem state associated with the AC involving the ground-state. Further details on the construction of these more general instances can be found in \ref{app:mult-opt}. For all the examples examined (where the problem parameters were chosen to produce a perturbative crossing with strong gap scaling) we observed a closing of the gap at the AC as well as the formation of an additional gap minimum between the instantaneous ground and first excited states - suggesting that the effects from the catalyst described in Section \ref{sec:closing_gaps} may be a more general phenomenon. An example spectrum with and without a catalyst is shown in Figure \ref{fig:mult-local-opt}.

We note however that the production of this additional gap minimum cannot generally be expected to result in a setting suitable to DQA. For instance, consider the example energy spectra presented in Figure \ref{fig:mult-local-opt} where we have introduced an additional local optimum. Here, the additional gap minimum results in a setting where the system is likely to evolve to the second excited state if the anneal is run diabatically. We note that if the anneal is run without a catalyst, the energy spectrum suggests that the system is likely to evolve to the first excited state - meaning that, in this case, the additional gap minimum not only fails to help the system reach the ground-state, but it actually reduces the quality of the solution found. We therefore suggest that, while the mechanism responsible for this additional gap minimum may be of interest for manipulating the gap spectrum, the changes to the gap spectrum we observe in this work are not, by themselves, useful for creating a gap spectrum suitable to DQA.

We note that the formation of the additional gap minimum that we observe in the strong scaling case is not intrinsically linked to the presence of a perturbative crossing in the original annealing spectrum. We demonstrate this in Figure \ref{fig:no-AC-case} which shows results for the gap spectrum and ground-state evolution when a catalyst is applied to the setting without an AC shown in Figure \ref{fig:no-cat-results}(a). The catalyst contains an XX-coupling within $G_0$ such that it couples $\ket{E_0}$ to a lower energy state than it couples $\ket{E_1}$ to. We suggest that the formation of the new gap minimum may relate to the existence of a $J_{xx}$ value for which both the problem ground and first excited state components of the instantaneous ground-state vector have appreciable magnitudes in the vicinity of $s_-$ such that either component crossing zero results in a sharp change in the ground-state vector.

We argue that the reason such a value is observed for the setting without an AC as well as the strong AC setting (but not for the weak AC) relates to the magnitudes of key ground-state vector components in the original annealing spectrum. Effectively, what we are doing with the introduction of our catalyst is to guide the anneal away from a particular problem state, which has the effect of suppressing the magnitude of the corresponding vector component in the instantaneous ground-state. For the two settings with perturbative crossings that we examined this is the component corresponding to the problem first excited state and for the setting without a perturbative crossing this is the component corresponding to the problem ground-state. What the setting without an AC and the strong AC setting have in common is that the magnitude of the component being suppressed ($\bra{E_0(s)} E_{0}\rangle$ for the no AC case and $\bra{E_0(s)} E_{1}\rangle$ for the SGS AC case) approaches $1$ in the original annealing spectrum without a catalyst - as can be seen in Figures \ref{fig:no-cat-results}(a) and (b). Looking at Figure \ref{fig:no-cat-results}(c), we see that in the weak AC setting, the maximum magnitude reached by the component being suppressed ($\bra{E_0(s)} E_{1}\rangle$) is only around $0.45$. Thus, the problem state that our catalyst is guiding the anneal away from reaches greater levels of suppression for the same $J_{xx}$ values in this setting than it does in the other two. This goes hand in hand with our earlier observation that, unlike in the SGS setting, the AC is already significantly lifted in the WGS setting for the $J_{xx}$ value at which $s_\textrm{\scriptsize x} = s_-$.

It is not yet clear whether the behaviour we observe in these two settings will scale to larger systems. Regarding the gap enhancement in the weak AC setting, the optimal $J_{xx}$ appears to quickly approach a constant - see the inset in Figure \ref{fig:cat-gap-scaling}(a). This implies that a single XX-coupling applied at a constant strength can be used to lift a perturbative crossing - even as this single coupling becomes negligible with respect to the system it is being applied to. While this seems counter-intuitive, such phenomena are not unfamiliar in condensed matter physics, where a single impurity spin in a system can have a significant effect \cite{Eggert1992,Laflorencie2008}. That being said, the system sizes that we have looked at may not be sufficient to determine the true scaling behaviour. If the scaling trend does persist however, this would be very promising since the number of potential XX-couplings scales as the square of the system size. This means that if only one XX-coupling is required to remove a perturbative crossing, an appropriate coupling may be found in polynomial time using trial and error. 

This does not however imply that an algorithm that utilises these findings will allow QA to find the ground-state of the problem Hamiltonian in polynomial time. The way in which we scale our problem graph in this work is to increase the number of vertices while keeping the number of local optima constant. However, in a realistic setting, one would expect the number of local optima to scale with the problem size. Our results suggest that a single non-stoquastic XX-coupling may be sufficient to remove one perturbative crossing if it is chosen to couple the local optimum responsible for the perturbative crossing to another low energy state. But, in a setting with multiple local optima, several perturbative crossings may have to be removed before the minimum gap scaling between the ground and first excited state becomes polynomial. We have begun investigating whether these catalysts could be extended to include multiple couplings to target the different perturbative crossings in the spectrum - with our initial results suggesting that a similar reduction to the gap scaling can be achieved in this way. We suggest that an algorithmic approach similar to that in \cite{Dickson2012} could be developed, where couplings are introduced and their strengths adjusted according to the local optima returned by sub-exponential QA runs. It is plausible that, if such a process could produce a spectrum that allowed a sub-exponential annealing run to achieve a high GS fidelity, it may require exponentially many calls to the quantum annealer. Something else to consider is how this might work for more general graph structures. Our preliminary results for problems with multiple local optima are using the graphs outlined in \ref{app:mult-opt} for which all the local optima are well separated in Hamming weight. However, in more general settings, local optima of the MWIS problem may share vertices. It is possible that, under these conditions, introducing XX-couplings recursively in the way we propose may sometimes result in the kind of correlated double ACs seen in \cite{Choi2021}. It would be interesting to see if an algorithmic approach that naturally combined these two effects produced a path through the annealing spectrum that allowed the GS to be reached by a sub-exponential annealing run.

Returning to the SGS setting, it also remains to be seen whether or not there continues to exist a regime where we observe the closing gaps and where the catalyst does not reduce the gap scaling as we move to larger problems. By exploiting permutation symmetries in the problem \cite{Albash2019} we have been able to begin preliminary investigations into graphs with up to 70 vertices and have confirmed the existence of the additional closing gaps up to these system sizes. Something it may be beneficial to investigate is how the boundary between the two regimes changes with respect to different problem parameters with increasing system size. With regards to the existence of additional local optima, we have so far confirmed the behaviour we associated with the strong gap scaling for graphs with up to five local optima. Further numerical investigations, and a better understanding of what causes the closing gaps that we have observed, are required to establish under what conditions we will observe the contrasting effects examined in this paper, and so how our findings may translate to more realistic problem settings.

Finally, we note that while our investigation has made use of the MWIS problem, there is no reason to think these findings should not extend to other problem settings. Crucially, the motivation behind the catalyst, and our further discussion of its effects, have hinged on the structure of the problem spectrum - rather than specific features of the MWIS problem. We would therefore expect to see similar results for any setting where an XX-term can be introduced to couple a local optimum responsible for a perturbative crossing to another low energy state. Further to this, we suggest that catalysts containing higher order X couplings, or couplings of a different form (e.g: YY, XY etc...), could also be understood in this way - with the effect that they have on the annealing spectrum being tied to the couplings which they create between states in the problem spectrum. 

\section{Conclusion}

We have examined the effects of specific XX-catalysts on annealing spectra corresponding to small instances of the MWIS problem. In particular, we have examined how the effect of a targeted non-stoquastic XX-coupling, chosen to enhance the gap size at a perturbative crossing, differs depending on the nature of the AC that is present in the spectrum. We found that the response of the gap spectra to the introduction of the catalyst was highly sensitive to subtle changes in the encoding of the problem which affected the scaling of the gap at the AC with system size

More specifically, we found that for parameter settings that resulted in a AC with milder (but still exponential) gap scaling, the catalyst resulted in an enhancement of the gap minimum, and that if an optimised catalyst strength was used for each system size, the gap scaling could be significantly improved. We then applied the same catalysts to the case where we chose our problem parameters to produce an AC with a stronger gap scaling and found that, while some gap enhancement was possible, the catalyst was not able to provide the same improvement in scaling. In addition, we found that in this setting the catalyst could result in the closing of the gap at the AC, as well as the production of an additional gap minimum in the spectrum, if introduced with particular magnitudes. We determined that these magnitudes were the ones for which the $s$ value at which the gap minimum occurred was the same as that for which the sign of key ground-state vector components changed. 

These results suggest that the catalysts we examine here may be less successful in suppressing the gap scaling at an AC when the original gap scaling is less favourable. More generally, they show that small changes to the problem parameters can result in the same catalyst having strikingly different effects indicating that great care is required when designing a catalyst.

With regards to the creation of the new gap minimum, we found that the resultant spectrum allowed a higher GS fidelity to be reached for shorter run times as a result of diabatic transitions. We noted however that the introduction of the additional gap minimum could not generally be expected to produce such a spectrum and so was not, by itself, a strategy for creating spectra suitable for DQA. 

As well as helping the guide the development of QA algorithms, these findings may also have relevance for other NISQ-era quantum algorithms such as the Quantum Approximate Optimisation Algorithm (QAOA) \cite{farhi2014quantum} which can be thought of as a trotterised QA. Coherence times limit QAOA to shallow circuit depths in much the same way that individual annealing runs are limited to short duration. Hamiltonians that reduce the required run-time in QA may also be a useful tool in helping QAOA reach higher GS fidelities for smaller circuit depths.

\begin{acknowledgments}
We gratefully acknowledge Vicky Choi, Tameem Albash, Daniel O’Connor and Robert Banks for inspiring discussion and helpful comments. This work is supported by EPSRC, grant references EP/S021582/1 and EP/T001062/1.
\end{acknowledgments}

\bibliography{refs.bib}

\appendix

\section{MWIS problem parameters \label{app:prop-parameter-tuning}}

The MWIS problem Hamiltonian is given by 
\begin{equation}
\label{eq:Prob-Hamiltonian}
    H_p = \sum \limits_{i \in \{\textrm{\scriptsize vertices}\}} (c_i J_{zz} - 2w_i)\sigma^z_i + \sum \limits_{(i,j) \in \{\textrm{\scriptsize edges}\}} J_{zz} \sigma^z_i \sigma^z_j
\end{equation}
where $c_i$ is the number of edges connected to vertex $i$, $w_i$ is the weight on vertex $i$, and $J_{zz}$ is the edge penalty. As discussed in Section \ref{sec:graph}, our problem instances are on complete bipartite graphs where we label our two disconnected sub-graphs as $G_0$ and $G_1$ - as shown in Fig \ref{fig:problem-graphs}. We allocate a total weight of $1$ to $G_1$ and a total weight of $1+\delta W$ to $G_0$. This weight is split evenly between the vertices in each sub-graph. To get a clearer picture of what parameters we can change to define our problem instances within this structure, we can re-write equation \ref{eq:Prob-Hamiltonian} as
\begin{multline}
\label{eq:specific-problem}
    H_p = (n_1 J_{zz} - \frac{2(1+\delta W)}{n_0}) \sum \limits_{i \in G_0} \sigma^z_i \\ + (n_0 J_{zz} - \frac{2}{n_1})\sum \limits_{i \in G_1} \sigma^z_i \\ + J_{zz} \sum \limits_{i \in G_0} \sum \limits_{j \in G_1} \sigma^z_i \sigma^z_j,
\end{multline}
where $n_0$ and $n_1$ are the number of vertices in $G_0$ and $G_1$ respectively. For a particular problem graph defined by $n_0$ and $n_1$, the free parameters are then $\delta W$ and $J_{zz}$.

So that we do not change the energy scale of the problem Hamiltonian when we adjust the values of $\delta W$ and $J_{zz}$, but rather just the relative gaps between different energy levels, we normalise the parameters that go into equation \ref{eq:Prob-Hamiltonian}. For a set of un-normalised parameter values $(\delta W', J'_{zz})$ that define the problem instance, we first calculate the un-normalised vertex weights as $w'_i = (1+\delta W' )/n_0$ for $i \in G_0$ and $w'_i = 1/n_1$ for $i \in G_1$ - giving us our set of un-normalised parameters $(\{w'_i\}, J'_{zz})$. The normalised parameters are then obtained as
\begin{equation*}
    w_i = E_\textrm{\scriptsize scale} \times K \times w'_i,
\end{equation*}
\begin{equation*}
    J_{zz} = E_\textrm{\scriptsize scale} \times K \times J'_{zz}
\end{equation*}
where $E_\textrm{\scriptsize scale}$ sets the energy scale in relation to the driver and $K$ is a normalisation factor given by
\begin{equation*}
    K  =  \frac{n}{4(N_\textrm{\scriptsize edges} \times J'_{zz} - W)} = \frac{n_0 + n_1}{4(n_0 \times n_1 \times J'_{zz} - 1)}.
\end{equation*}
This normalisation factor has been obtained by using equation \ref{eq:specific-problem} to calculate the difference in energy between the ground and highest excited state.

We choose our problem parameters with reference to our five vertex instance ($n_0=2$ and $n_1=3$) since this is the instance for which we present data for the whole anneal rather than just (e.g.) the minimum gap value. Our different parameter sets $(\delta W', J'_{zz})$ are chosen by first selecting $\delta W'$ and then, through trial and error, adjusting the \textit{un-normalised} $J'_{zz}$ so that in the annealing spectrum to corresponding to the \textit{normalised} problem Hamiltonian, $s_\textrm{\scriptsize x}=0.9$ for our $5$-vertex example. Note that the values for $\Delta W$ and $J_{zz}$ quoted in the body of this work are in fact the un-normalised values $\Delta W'$ and $J_{zz}'$. 

\section{Perturbative introduction of the catalyst \label{app:pert_argument}}

\begin{figure*}[]
    \centering
    \includegraphics[scale=0.33]{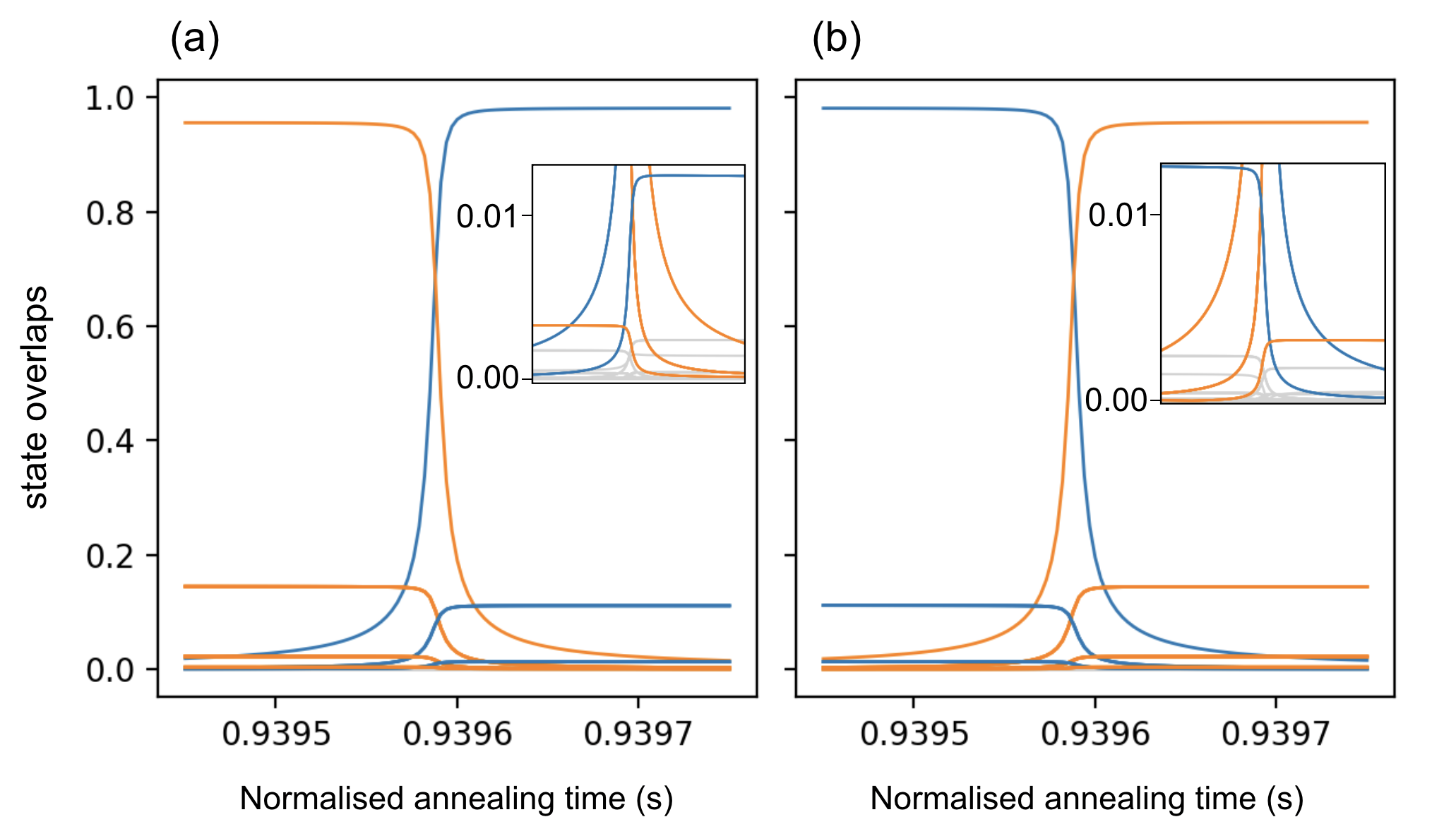}
    \caption{ Evolution of the instantaneous ground-state (a) and first excited state (b) around the location of the perturbative crossing in an anneal corresponding to a problem instance with $G_0 = 3$, $G_1 = 4$, $\Delta W = 0.01$ and $J_{zz} = 5.33$. The instantaneous states are represented in terms of their overlaps with the problem states and these are colour coded with respect to which of the sets introduced in \ref{app:pert_argument} they are in. Overlaps with problem states in $P_{0}^\textrm{\scriptsize ext}$ and $P_{1}^\textrm{\scriptsize ext}$ and plotted in blue and orange respectively and overlaps with problem states that are in neither are plotted in light grey. The insets show the same data over a smaller range of the state overlaps so that the overlaps with smaller contributions are visible. }
    \label{fig:crossing_example}
\end{figure*}

\begin{figure*}[]
    \centering
    \includegraphics[scale=0.27]{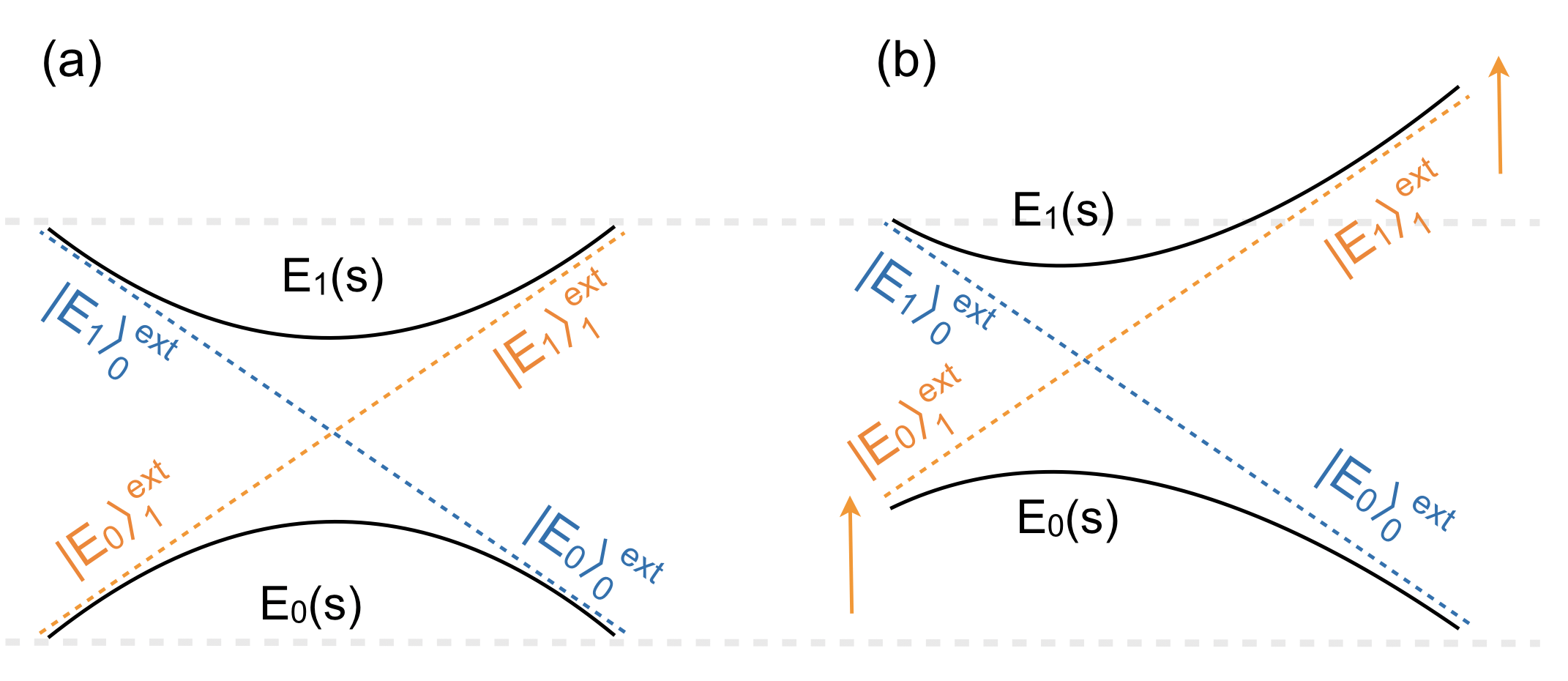}
    \caption{ Cartoons illustrating the arguments made in \ref{app:pert_argument} regarding how the introduction of our choice of catalyst would perturb the instantaneous energy levels around the perturbative crossing. (a) depicts an avoided level crossing in which the ground-state goes from being dominated by the set of problem states $P_{1}^\textrm{\scriptsize ext}$ to the set of problem states $P_{0}^\textrm{\scriptsize ext}$ while the first excited state goes from being dominated by $P_{0}^\textrm{\scriptsize ext}$ to $P_{1}^\textrm{\scriptsize ext}$. The two coloured dotted lines illustrate the exchange in problem states between the instantaneous ground and first excited states and the solid black lines illustrate the energies of the two instantaneous states. (b) then depicts the effect of the catalyst as argued in \ref{app:pert_argument}. That is, there is no effect on a state consisting of $P_{0}^\textrm{\scriptsize ext}$ while a state consisting of $P_{1}^\textrm{\scriptsize ext}$ will experience an increase in energy. The cartoon illustrates why we expect this to move the crossing earlier in the anneal. }
    \label{fig:pert-cat-cartoon}
\end{figure*}

This section follows a perturbative argument made in \cite{Choi2021} which offers some insight into why the catalysts used in this work result in gap enhancement at the perturbative crossings in our systems. Unlike the perturbative argument outlined in Section \ref{sec:graph}, in which $H_p$ was taken to be the initial Hamiltonian, here we consider the effect of introducing the catalyst as a perturbation to the total Hamiltonian for a given $s$. That is, we take the initial Hamiltonian to be $H(s) = (1-s)H_d + sH_p$ and then write our perturbed Hamiltonian as
\begin{equation*}
    H(s,\mu) = H(s) + \mu H_c.
\end{equation*}
To first order, the perturbed energies are then
\begin{multline}
    \label{eq:cat-pert}
    E_i(s,\mu) = E_i(s) + \mu \bra{E_i(s)}H_c\ket{E_i(s)} \\ = E_i(s) + 2 \mu J_{xx} \sum \limits_{(j,k) \in C_{xx}} \braket{E_i(s)|E_j}\braket{E_i(s)|E_k}
\end{multline}
where we have used $C_{xx}$ to denote the set of pairs of problem states that are coupled by the single XX-coupling included in the catalyst.

We are then interested in what happens to the instantaneous ground and first excited states ($i=0,1$) around the point of the perturbative crossing, $s_\textrm{\scriptsize x}$. In order to use equation \ref{eq:cat-pert} to gain some insight as to the effect of the catalyst we can consider the perturbative argument from Section \ref{sec:graph} to reason what problem states will dominate the two instantaneous states \textit{prior} to the introduction of the catalyst - i.e: for $H(s,\mu = 0)$. We know that for $s>s_\textrm{\scriptsize x}$, $\ket{E_0}$ will make up a large part of $\ket{E_0(s)}$ and that $\ket{E_1}$ will make up a large part of $\ket{E_1(s)}$. As for what other problem states will have a reasonable presence in $\ket{E_{0,1}(s)}$, we know from perturbation theory that the introduction of the driver will introduce states with a weight that depends on their energy difference and Hamming distance from the state being perturbed - i.e: from $\ket{E_{0,1}}$. We thus reason that for $s>s_\textrm{\scriptsize x}$, $\ket{E_{0}(s)}$ is occupied by problem states corresponding to sub-sets of $G_0$ and that $\ket{E_{1}(s)}$ is occupied by problem states corresponding to sub-sets of $G_1$. That is, they are occupied by states corresponding to independent sets (i.e: low energy states such that the energy differences between them and $E_{0,1}$ are small) that are closest in Hamming weight to the two respective problem states. We denote these sets of problem states as $P_{0,1}^\textrm{\scriptsize ext}$. 

We wish to examine the effect of the catalyst at points in the anneal shortly before and after the AC. Following the notation in \cite{Choi2021}, we denote these points as $s_\textrm{\scriptsize x}^-$ and $s_\textrm{\scriptsize x}^+$ respectively. From the above, we can write
\begin{equation*}
    \ket{E_0(s_\textrm{\scriptsize x}^+)} \approx \sum \limits_{i \in P_0^\textrm{\scriptsize ext}} \braket{E_0(s_\textrm{\scriptsize x}^+)|E_i} \ket{E_i} \equiv \ket{E_0}_0^\textrm{\scriptsize ext}
\end{equation*}
where we are not interested in the specific values of $\braket{E_0(s)|E_i}$. Note that the subscript inside the ket refers to the instantaneous ground-state while the subscript outside the ket refers to the set of states $P_{0}^\textrm{\scriptsize ext}$. Similarly, we can write
\begin{equation*}
    \ket{E_1(s_\textrm{\scriptsize x}^+)} \approx \ket{E_1}_1^\textrm{\scriptsize ext}.
\end{equation*}
At the AC, the two instantaneous states exchange their properties and so we can further write
\begin{equation*}
    \ket{E_0(s_\textrm{\scriptsize x}^-)} \approx \ket{E_0}_1^\textrm{\scriptsize ext},
\end{equation*}
\begin{equation*}
    \ket{E_1(s_\textrm{\scriptsize x}^-)} \approx \ket{E_1}_0^\textrm{\scriptsize ext}.
\end{equation*}
We support our assumptions about the makeup of the instantaneous states around the AC with numerical results shown in Figure \ref{fig:crossing_example}. These results correspond to a problem instance with $G_0 = 3$, $G_1 = 4$, $\Delta W = 0.01$ and $J_{zz} = 5.33$. Figures \ref{fig:crossing_example}(a) and (b) show the evolution of the instantaneous ground and first excited state respectively around the location of the perturbative crossing. Plotted in blue are the overlaps of the instantaneous states with the problem states in $P_{0}^\textrm{\scriptsize ext}$ and in orange we show the overlaps with the problem states in $P_{1}^\textrm{\scriptsize ext}$. The rest are plotted in grey. These results are in agreement with our predictions.

We now consider what the perturbation in equation \ref{eq:cat-pert} looks like when the state being perturbed is  $\ket{E_{0,1}}_0^\textrm{\scriptsize ext}$ or  $\ket{E_{0,1}}_1^\textrm{\scriptsize ext}$. Due to the fact that the XX-coupling we introduce is between two spins in $G_1$, all the states in $P_0^\textrm{\scriptsize ext}$ are coupled by the catalyst to states corresponding to dependent sets. That is there are no pairs of states in $C_{xx}$ for which both states have a significant presence in $\ket{E_{0,1}}_0^\textrm{\scriptsize ext}$. As a result, $\bra{E_{0,1}}_0^\textrm{\scriptsize ext}H_c\ket{E_{0,1}}_0^\textrm{\scriptsize ext} = 0$. Thus, $E_0(s_\textrm{\scriptsize x}^+,\mu) \approx E_0(s_\textrm{\scriptsize x}^+)$ and $E_1(s_\textrm{\scriptsize x}^-,\mu) \approx E_1(s_\textrm{\scriptsize x}^-)$.

On the other hand, there are pairs of states in $C_{xx}$ for which both states have a significant presence in $\ket{E_{0,1}}_1^\textrm{\scriptsize ext}$. Let us first consider what this means for $E_0(s_\textrm{\scriptsize x}^-,\mu)$. Because $H(s)$ is stoquastic, the ground-state vector components, $\braket{E_0(s)|E_i}$, must all be positive meaning that all the terms in the sum in equation \ref{eq:cat-pert} are also positive. Thus, for $J_{xx}>0$ (as is the case in our setting), $E_0(s_\textrm{\scriptsize x}^-,\mu) > E_0(s_\textrm{\scriptsize x}^-)$. We now consider the perturbed energy $E_1(s_\textrm{\scriptsize x}^-,\mu)$. In this case, the individual vector components cannot be guaranteed to be positive since the first excited state can have both positive and negative vector components. However we can use an observation made in \cite{Choi2021}, regarding the behaviour of the vector components of two instantaneous states at an AC, to argue that $sign(\braket{E_0(s)|E_i}\braket{E_0(s)|E_j}) = sign(\braket{E_1(s)|E_i}\braket{E_1(s)|E_j})$. Thus, we can also say that all the terms in the sum will be positive and that $E_1(s_\textrm{\scriptsize x}^+,\mu) > E_1(s_\textrm{\scriptsize x}^+)$.

In order to make a rigorous case for this enhancing the minimum gap size at the AC, further ideas from \cite{Choi2021} must be introduced. However, some intuition as to why the gap size increases can be had by considering the cartoons shown in Figure \ref{fig:pert-cat-cartoon}. Figure \ref{fig:pert-cat-cartoon}(a) depicts the kind of AC described above where the instantaneous ground-state goes from $\ket{E_{0}}_1^\textrm{\scriptsize ext}$ to $\ket{E_{0}}_0^\textrm{\scriptsize ext}$ while the instantaneous first excited state goes from $\ket{E_{1}}_0^\textrm{\scriptsize ext}$ to $\ket{E_{1}}_1^\textrm{\scriptsize ext}$. The dashed lines illustrate the $\ket{E_{0,1}}_{0,1}^\textrm{\scriptsize ext}$ states (which can also be thought of as the perturbed states obtained when introducing the driver perturbatively to the problem Hamiltonian as in Section \ref{sec:graph}) and the solid black lines show the two instantaneous energy levels. Above we have argued that $\ket{E_{0,1}}_1^\textrm{\scriptsize ext}$ is increased by the introduction of the catalyst while $\ket{E_{0,1}}_0^\textrm{\scriptsize ext}$ is unchanged. Figure \ref{fig:pert-cat-cartoon}(b) depicts the effect of introducing the catalyst. We can see that the result is to move the AC to a lower value of $s_\textrm{\scriptsize x}$. Thus the strength of the driver Hamiltonian will be greater at the point of the crossing and the vectors will be more mixed - thus resulting in a greater overlap between the instantaneous ground-state before and after the crossing and so a larger gap size. We note that the decrease in $s_\textrm{\scriptsize x}$ is also consistent with our numerical results in Section \ref{sec:results}.

\section{Diabatic Anneal on the $5$-vertex Graph  \label{app:DQA}}

\begin{figure}[]
    \centering
    \includegraphics[scale=0.28]{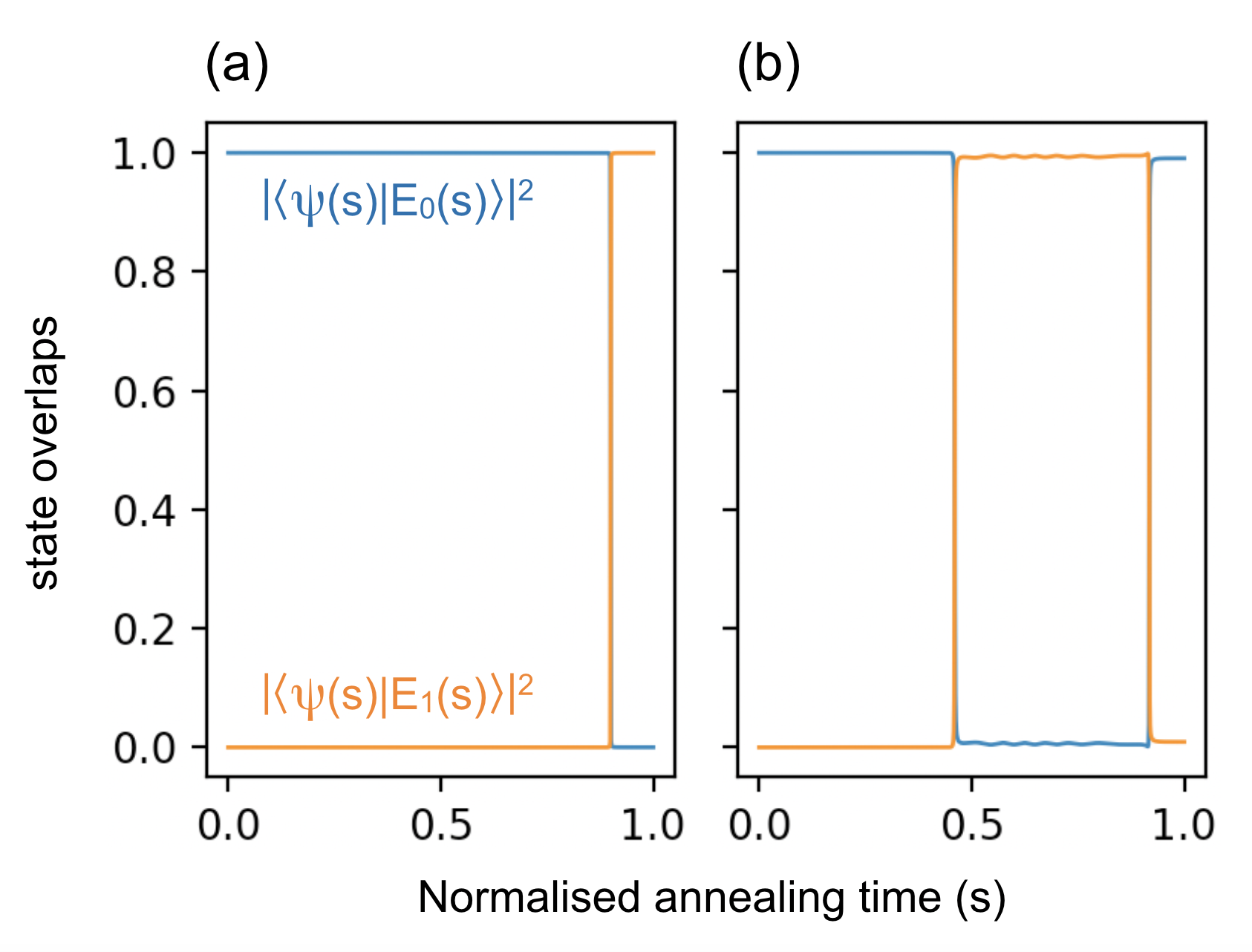}
    \caption{ Results showing the dynamics of an anneal to a problem graph with $n_0=2$, $n_1=3$, $\delta W=0.01$ and $J_{zz}=5.33$ in a closed system setting. This evolution is presented in terms of the system's overlap with the instantaneous ground and first excited states in blue and orange respectively. The total annealing time used is $T_\textrm{\scriptsize  anneal}=1000$ in comparison to the local driver fields being introduced with a magnitude of $1$. (a) shows the results without a catalyst and (b) shows the results using a catalyst with $J_{xx}=1.92$. }
    \label{fig:dynamics}
\end{figure}

\begin{figure*}[]
    \centering
    \includegraphics[scale=0.3]{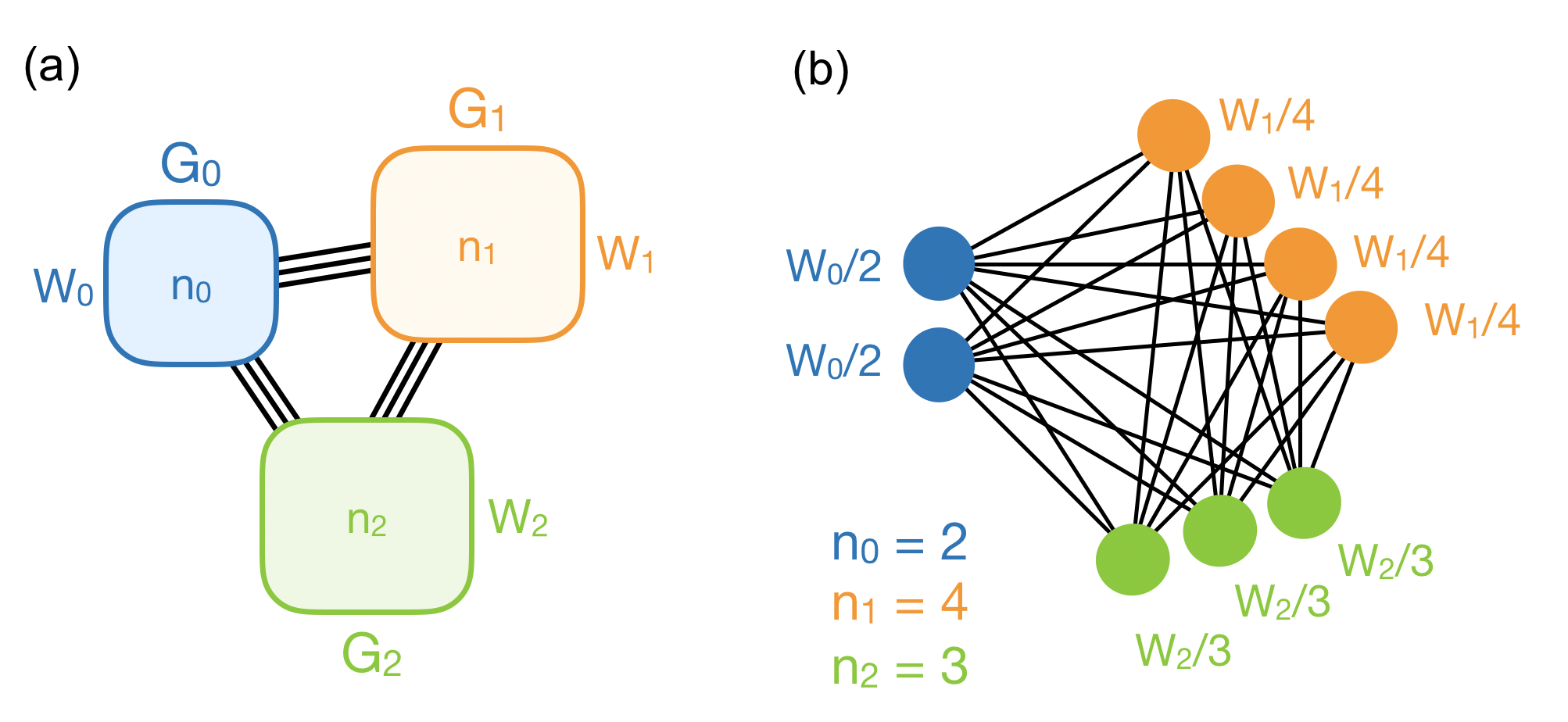}
    \caption{ An illustration of tri-partite graph constructed as described in appendix \ref{app:mult-opt} is shown in (a). An example with $n_0=2$, $n_1=4$ and $n_2=3$ is shown in (b). This is the example corresponding to the annealing spectra in Figure \ref{fig:mult-local-opt}. }
    \label{fig:mult-local-opt-graphs}
\end{figure*}

We present here numerical results for the dynamics of an anneal on the $5$-vertex problem instance in the SGS setting. These results are obtained using closed system spin models and the evolution of the system is presented in terms of its overlap with the instantaneous ground and first excited states. Since we have not specified a true energy scale we can cannot discuss the evolution with respect to an actual annealing time but only in relation to the magnitude with which we introduce our Hamiltonian. 

The annealing time used in these simulations was $1000$, in relation to the local driver fields being introduced with a magnitude of $1$. Figure \ref{fig:dynamics}(a) shows the evolution without a catalyst and we see that for this annealing time the system has a negligible overlap with the problem ground-state at the end of the anneal after transitioning into the first excited state at the avoided level crossing. Figure \ref{fig:dynamics}(b) shows the results when a catalyst is introduced with $J_{xx}=1.92$, such that another small gap in the spectrum is produced. We observe that in this case the system ends the anneal with a near unity overlap with the problem ground-state as a result of transitioning into the first excited state at the first small gap and then back into the ground-state at the second. 

We stress that these results are for a very specific system where only one perturbative crossing is present. The results presented in Figure \ref{fig:dynamics} show that the spectrum in Figure \ref{fig:new-gap-min}(c) does provide a diabatic path that allows the GS to be reached for shorter run times. However, as we discuss in Section \ref{sec:discussion}, we do not expect the creation of the additional gap minimum to always produce such a spectrum.

\section{Introducing additional local optima  \label{app:mult-opt}}

We describe here how our problem graph can be generalised to include additional local optima. This can be done straightforwardly by adding further sub-graphs for each additional local optima to produce a complete k-partite graph. That is we have k sub-graphs where each vertex in every sub-graph is connected to every vertex in every other sub-graph and there are no connections within the sub-graphs themselves. Each local optimum then corresponds to picking all the vertices from one of the sub-graphs. As with our bipartite graphs, each sub-graph $G_a$ is given $n_a$ vertices which determines the energy spectrum of the neighbourhood of the corresponding problem state. It is also given a total weight, $W_a$, which is shared out equally between its vertices such that each vertex in $G_a$ has a weight of $W_a/n_a$. A general tri-partite graph is shown in Figure \ref{fig:mult-local-opt-graphs}(a) and a specific example is shown in Figure \ref{fig:mult-local-opt-graphs}(b). However, more sub-graphs can be added.

\end{document}